\documentclass[12pt,a4paper]{article}
\usepackage{latexsym}
\usepackage{graphicx}
\usepackage{amsmath, amssymb}
\usepackage{mathrsfs}
\usepackage{natbib}
\usepackage{hyperref}
\usepackage{enumerate}

\usepackage{color}

\allowdisplaybreaks

\usepackage[latin1]{inputenc}
\usepackage[T1]{fontenc}

\newtheorem{theorem}{Theorem}[section]
\newtheorem{lemma}[theorem]{Lemma}
\newtheorem{definition}[theorem]{Definition}
\newtheorem{proposition}[theorem]{Proposition}
\newtheorem{corollary}[theorem]{Corollary}
\newtheorem{remark}[theorem]{Remark}
\newtheorem{example}[theorem]{Example}

\numberwithin{equation}{section}

\begin{document}
\newcommand{\FR}[1]{{\color{red} #1}}

\newcommand{\mathbbold}[1]{{ #1}}





\title{Optimal consumption and portfolio choice with ambiguity}

\author{Qian Lin \footnote{{\it Email address}:
Qian.Lin@uni-bielefeld.de}, Frank Riedel
\\
{\small  Center for Mathematical Economics, Bielefeld University, Germany}\\
 } \maketitle

\begin{abstract}\vspace{-0.2cm}
We consider  optimal consumption and portfolio choice
 in the presence of Knightian uncertainty in continuous-time. We embed the problem into the new framework of stochastic
calculus for such settings, dealing in particular with the issue of non--equivalent multiple priors.  We   solve the problem completely by identifying the worst--case measure.
Our setup also allows to consider interest rate uncertainty; we show that under some robust parameter constellations, the investor optimally puts all his wealth into the asset market, and does not save or borrow at all.
 \end{abstract}

\newpage

\section{Introduction}

The optimal way to invest and consume one's wealth belongs to the basic questions of finance. The standard textbook answer uses Merton's (\citeyear{Merton69})  solution within the framework of the geometric Brownian motion model for risky assets. In this paper, we generalize this fundamental model to allow for Knightian uncertainty about asset and interest rate dynamics and study the consequences for ambiguity--averse investors.

In continuous time, Knightian uncertainty leads to some subtle issues. Uncertainty about volatility, as well as uncertainty about the short rate, requires the use of singular probability measures, a curious, but
in an ambiguous world natural fact. The investor is cautious and presumes that nature has unpleasant
surprises. In particular, the
volatility of risky assets can take surprising paths, within certain limits.

 Fortunately, the last years have seen the development of a new stochastic calculus\footnote{\cite{DenisMartini06} developed such a framework for studying the pricing of options under Knightian Uncertainty; \cite{Peng07} develops the whole theory of stochastic calculus from scratch under Knightian uncertainty. } that extends the omnipresent  It\^{o}--calculus to such multiple prior models . The rationality of such an approach as well as the consequences for utility theory and equilibrium asset pricing have recently been discussed at length by \cite{EpsteinJi11}.   We show here how to embed the classic Merton--Samuelson model into this new framework. The new framework has the advantage that it allows to use essentially well-known martingale arguments to establish optimality of candidate policies, just as in the classic case.

While explicit results are difficult to obtain under Knightian uncertainty in general, we are here able to solve completely  the ambiguity--averse investor's optimal consumption--portfolio problem. In a first step, we derive the extension to Knightian uncertainty for the classic Hamilton--Jacobi--Bellman equation. A closer analysis of that equation leads to a conjecture for the worst--case measure. We then verify that the ambiguity--averse investor behaves as a classic expected--utility maximizer under the worst--case measure by using the new techniques.
The existence of a worst case measure immediately yields a maxmin result: the value function under ambiguity is the lower envelope of the value functions under expected utility.

Ambiguity leads to different predictions for optimal portfolios and consumption plans. As in simple static models, high ambiguity about the mean return of the uncertain asset leads to non--participation in the asset market. As far as volatility is concerned, we show that our risk-- and ambiguity--averse investor always uses the maximal possible volatility to determine the optimal policy.
A more surprising and, as far as we know, new result emerges when we take interest rate uncertainty into account: for robust parameter sets, the investor puts all his wealth into the asset market when interest rate uncertainty is sufficiently high, a phenomenon that we have observed in the aftermath of the recent financial crisis. When interest rates are low, and sufficiently high ambiguity is perceived by investors, they prefer to put all capital  into assets only. Both saving and borrowing are considered to be too uncertain to be worthwhile activities.

The problem of Knightian or model uncertainty  has recently attracted a great deal of attention, both in practice, as the
 sensitivity of many financial decisions with respect to questionable probabilistic assumptions became clear, and in theory, where an
  extensive theory of decision making and risk measurement under uncertainty has been developed.  \cite{GilboaSchmeidler89}   lay the foundation for a new approach to
decisions under Knightian uncertainty by weakening the strong
independence axiom or sure thing principle used previously by \cite{Savage54} and \cite{AnscombeAumann63} to justify (subjective)
expected utility. The models are closely related to monetary risk measures (\cite{Artzneretal99}).  Subsequently, the theory has been generalized to
variational preferences (\cite{MaccheroniMarinacciRustichini06}, \cite{FoellmerSchied02}) and dynamic time-consistent models (\cite{EpsteinSchneider03}, \cite{Maccheronietal06}, \cite{Riedel04}).

The pioneering results of \cite{Samuelson69} and \cite{Merton69} laid the foundation for a huge literature. As mean return,  volatility, and interest rates are constants in the basic model, the consequences of having stochastic, time-varying dynamics for these parameters have been studied in great detail. Mean--reverting drift (or ``predictable returns'')  , stochastic volatility models  and models with stochastic term structures have been studied in detail. These models all work under the expected utility paradigm as they assume a known distribution for the parameters. In the same vein, one can also study incomplete information models where the investor updates his initial belief about some unknown parameter\footnote{\cite{Barberis00} studies mean--reverting returns and estimation errors. \cite{ChackoViceira05} study stochastic volatility. See \cite{Liu07} for a recent general approach with stochastic interest rates and volatilities.
\cite{SchroderSkiadas02} }. In contrast to these Bayesian models, we focus here on the recent Knightian approach where the investor takes a pessimistic, maxmin view of the world concerning the parameters of her model. One can also relax the time--additive structure of the intertemporal utility function as in recursive utility models \cite{DuffieEpstein92}, Hindy--Huang--Kreps models (\cite{HindyHuang92}, \cite{BankRiedel01b}), or allow for trading constraints and transaction costs, topics that are outside the scope of this paper.

The Knightian approach is closely related to model uncertainty, or robustness considerations in the spirit of  \cite{Andersonetal03}.  For example, \cite{TrojaniVanini02}
 and \cite{Maenhout04}   study the robust
portfolio choice problem with drift ambiguity. Drift ambiguity in continuous time is also discussed in
 \cite{EpsteinChen02}, \cite{Miao09} , \cite{Schied05}, \cite{Schied08}, \cite{Liu10}, \cite{Liu11} among others. \cite{FoellmerSchiedWeber09} survey this literature.
 In these papers, a reference measure to which all priors are equivalent is used,  in contrast to our approach. In particular, one cannot discuss volatility uncertainty within these models. 

The paper is setup as follows. The next section formulates the Merton model under Knightian uncertainty within the new framework. The following section derives derives optimal consumption and portfolio rules for ambiguity-averse investors with fixed interest rates. Section 4 then generalizes to ambiguous interest rates. An appendix collects the relevant information about the new stochastic calculus.

\section{The Samuelson-Merton Model under Knightian Uncertainty}\label{2}

The standard workhorse for asset pricing in continuous time has been
proposed by Samuelson (1969) and Merton (1971); they work with a
safe asset, or bond, with deterministic dynamics
$$ dP_t = r P_t dt$$ for a known interest rate $r$ and a risky asset $S$ that satisfies
$$dS_t = \mu S_t dt + \sigma S_t dB_t$$ for a Brownian motion $B$ and known drift and volatility parameters $\mu$ and $\sigma$.

This basic model has been extended in many forms, of course, as we discussed in the introduction.. Here, we show how to treat the optimal portfolio-consumption choice problem for an investor who does not know the specific parameters nor their probability laws. We thus have Knightian uncertainty in the sense that the distribution of the unknown parameters is not known. However, the investor is willing to work with (or knows) certain bounds for the relevant parameters; she is ambiguity--averse and aims to find a policy that is robust to such parameter uncertainty in the sense that it is optimal even against a malevolent nature.

As far as modeling is concerned, our new approach shows how to embed the Samuelson--Merton model in an extended stochastic calculus framework; the advantage of that model is that it allows to use the well-known It\^{o}-calculus-type martingale arguments to solve the problem.

Technically, we will replace the standard Brownian motion $B$ by a so--called $G$--Brownian motion that we denote by the same symbol $B$. A $G$--Brownian motion is a diffusion with unknown volatility process. It shares many properties with the known Brownian motion of classic calculus; its quadratic variation $\langle B\rangle$, however, is not equal to expired time $t$; only  estimates of the form  $$\langle B\rangle_t \in \left[\underline{\sigma}^2, \overline{\sigma}^2\right]$$ for some volatility bounds $0<\underline{\sigma}\le  \overline{\sigma}$ are given. The classical model is recovered for $\underline{\sigma}=\overline{\sigma}$.

We will replace the drift term $\mu dt$ by an ambiguous term $db_t$ where $b$ is a process of bounded variation that allows for any drift between two bounds $[\underline{\mu}, \overline{\mu}]$.
Our new model for what we now call the uncertain, rather than the risky asset reads as
$$dS_t = S_t db_t + S_t dB_t \,.$$
We will write $dR_t = db_t + dB_t$ for the return process in the sequel. It exhibits mean and volatility uncertainty.

The riskless asset is standard, with  price dynamics
$$ dP_t = r P_t dt,$$ where
$r$ is the constant interest rate.  One can allow for interest rate
uncertainty as well as we show in   Section \ref{section4}. For the
moment, the results are more transparent when we keep
 the idealized assumption of constant and known interest rates.

The investors beliefs are summarized then by a set of priors of the
form $P^{\mu,\sigma}$ where $\mu$ and $\sigma$ are (progressively
measurable) stochastic processes. Under a prior $P^{\mu,\sigma}$,
the uncertain asset has drift $\mu$ and volatility $\sigma$, and the
short
 rate is equal to $r$. Note that for different volatility or short rate specifications,
  the priors are mutually singular to each other. The investor does not fix the null sets ex ante;
  under model uncertainty, one needs to reduce the number of ``impossible'' events.
  Only events that are null under all possible priors can be considered to be negligible.
  Technically, these events are called polar; if an event has probability one under all priors,
  we say it happens quasi-surely.

In the appendix, we describe the mathematical construction of the
set of priors $\mathcal {P}$ and the corresponding sublinear and
superlinear expectation for the more general multi-dimensional case
(using Shige Peng's method) in more detail. It can be skipped at
first reading.

\subsection{Asset Prices}

Let $B$ be a $G$-Brownian motion with volatility bounds
$[\underline{\sigma}^2, \overline{\sigma}^2]$. Let $b$  be two
maximally distributed increasing process with drift bounds
$[\underline{\mu},\overline{\mu}]$.

 The uncertain asset prices evolve as
$$dS_t = S_t dR_t , S_0=1 $$ where the  return dynamics satisfy
\begin{eqnarray*}
dR_{t}=db_{t} +dB_{t}\,.
\end{eqnarray*}
The locally riskless bond evolves as
$$ dP_t =P_t r dt , P_0=1\,.$$

\subsection{Consumption and Trading Opportunities}

The investor chooses a portfolio strategy $\pi$ and a consumption plan $c$.
Uncertainty reduces the set of possible consumption plans and trading strategies an investor might choose. This reflects the economic incompleteness of markets that uncertainty can bring. As in the classic case, we want to give precise meaning to the intertemporal budget constraint
\begin{align}
      dX_{t} & =
 X_t\pi_t^{T} dR_t + (1-\pi_t^{T}\mathbbold{1}) X_t r dt - c_t dt\nonumber \\
&=
rX_{t}(1-\pi_{t}^{T}\mathbbold{1})dt+X_{t}\pi_{t}^{T}db_{t}-c_{t}dt+X_{t}\pi^{T}_{t}AdB_{t}\,.
\end{align}
In order to do so, we have to introduce suitable restrictions, intuitively speaking, to make the stochastic differential equation meaningful under all priors simultaneously.

In order to give the definitions of a consumption plan and portfolio
choices precisely, we introduce  spaces of random variables and
stochastic processes, which are different from the classical case, technically speaking,
because of the nonequivalence of the priors.

We denote by
$\Omega=C_{0}^{2d}(\mathbb{R}^{+})$ the space of all
$\mathbb{R}^{2d}$-valued continuous path
$(\omega_{t})_{t\in\mathbb{R}^{+}}$ with $w_{0}=0$, equipped with
the topology generated by the uniform convergence on compacts.

 We let $L^{2}(\Omega)$
be the completion of the set of all bounded and continuous functions
on $\Omega$ under the norm
$\parallel\xi\parallel=\mathbb{\hat{E}}[|\xi|^{2}]^{\frac{1}{2}}:=\sup\limits_{P\in
\mathcal {P}}E_{P}[|\xi|^{2}]^{\frac{1}{2}}$. For $t\in [0,T]$, we
define the following space
\begin{eqnarray*}
L_{ip}(\Omega_{t})=\Big\{\varphi(\omega_{{t_{1}}}
\cdots, \omega_{{t_{m}}}) \ | \ \ m\in \mathbb{N}, \ t_{1}, \cdots, t_{m}\in [0,t], \\
\ \text{for all bounded function}\ \varphi\Big\}.
\end{eqnarray*}

 For $p\geq 1$, we now consider the process $\eta$ of the following form:
\begin{eqnarray*}
\eta=\sum\limits_{j=0}^{n-1}\xi_{j}1_{[t_{j},t_{j+1})},
\end{eqnarray*}
where $0=t_{0}<t_{1}<\cdots<t_{n}=T$, and $\xi_{j}\in
L_{ip}(\Omega_{t_{j}}), j=0,\cdots, n-1,$ We denote the set of the
above processes $M^{p, 0}$. And the norm in $M^{p, 0}$ is defined by
$$\parallel \eta
\parallel_{p}=\bigg(\mathbb{\hat{E}}\Big[\int_{0}^{T}|\eta_{t}|^{p}dt\Big]\bigg)^{\frac{1}{p}}=\bigg(\mathbb{\hat{E}}
\Big[\sum\limits_{j=0}^{n-1}|\xi_{t_{j}}|^{p}(t_{j+1}-t_{j})\Big]\bigg)^{\frac{1}{p}}.
$$
Finally, we denote by $M^{p}$ the completion of $M^{p, 0}$ under the
above norm.

The investor chooses a consumption plan $c$, a nonnegative
stochastic process  such that  $c\in M^{1}$. Also the investor can
choose the fraction of wealth $\pi^{i}_{t}$ invested in the $i$
risky asset, and the fraction of wealth
$1-\sum_{i=1}^{d}\pi^{i}_{t}$ invested in the riskless asset.

The wealth of the investor with some initial endowment $x_{0}> 0$ and portfolio--consumption policy $(\pi, c)$ is
given by
\begin{align}\label{p3}
      dX_{t} & =
 X_t\pi_t^{T} dR_t + (1-\pi_t^{T}\mathbbold{1}) X_t r dt - c_t dt\nonumber \\
&=
rX_{t}(1-\pi_{t}^{T}\mathbbold{1})dt+X_{t}\pi_{t}^{T}db_{t}-c_{t}dt+X_{t}\pi^{T}_{t}AdB_{t},
\end{align}
where $\pi=(\pi^{1},\cdots,\pi^{d})^{T}$ is the trading strategy,
and $\mathbbold{1}=(1,\cdots,1)^{T}$.  The consumption and portfolio
processes pair $(\pi, c)$ is admissible if $X_{t}\geq 0, t\in
[0,T]$,  $c\in M^{1}$ and  $\pi\in M^{2}$.

We denote by $\Pi$ the set of all such admissible $\pi$ taking value
in $\mathcal {B}=(-\infty,+\infty)$.  Also we denote by $C$ the set
of all such admissible $c$.

\subsection{Utility}

The investor is ambiguity--averse and maximizes the minimal expected
utility over his set of priors. For any random variable $X$ on
$(\Omega,\mathcal {F}_{T})$, we denote by $$ \underline{\mathbb{E}}
X :=\inf_{P \in {\mathcal P}} E_{P} X $$
the lowest expected value of an uncertain outcome $X$.

 The investor's utility of consuming $c \in M^1$ and bequesting a terminal wealth $X_T $ is
\begin{eqnarray*}
U(c,X)= \mathbb{\underline{E}}[\int_{0}^{T} u(s,
c_{s})ds+\Phi(T,X_T)],
\end{eqnarray*}
the utility function $u$ and the bequest function $\Phi$  are
strictly increasing, concave and  differentiable with respect to $c$
and $X$, respectively. We further suppose that $u$ and $\Phi$ are
$C^{1,3}$ and $C^{2}$, respectively. Furthermore, we suppose that
the marginal utility is infinite at zero:
\begin{eqnarray*}
 \lim\limits_{c\rightarrow 0}\dfrac{\partial}{\partial
 c}u(t,x)=\infty.
\end{eqnarray*}

We define the value function:
\begin{eqnarray*}
V(x_{0})=\sup\limits_{(\pi, c)\in\Pi\times C}  U(c, X),
\end{eqnarray*}
that is the indirect utility function defined over portfolio and
initial wealth. In the appendix, we prove that
 $V(x_{0})$  is increasing and concave in  $x$.

\section{Optimal consumption and portfolio choice with ambiguity}\label{section3}

\subsection {The Robust Dynamic Programming Principle}

We quickly recap the classical dynamic programming approach put
forward by Merton in one dimension. When $v(t,x)$ denotes the value
 function at time $t$ with wealth $x$, the dynamic programming principle states, informally, that
$$ v(t,X_t) \simeq \max_{\pi,c} u(t,c_t) + E [ v(t+\Delta t, X_{t+\Delta t} | {\cal F}_t ] $$
which, according to the usual rules of It\^{o} calculus, leads to
 the typical Hamilton-Jacobi--Bellman equation

$$\sup_{\pi, c} \Big\{ u(t,c) - c v_x(t,x) + \pi x (\mu-r) v_x (t,x)+ \frac{1}{2} \pi^2 x^2 \sigma^2 v_{xx} (t,x)\Big\} = 0 .$$

It reflects the usual martingale principle: for all admissible policies $(\pi, c)$ with wealth process $X$, the sum of indirect utility and past consumption utility
$$ v(t,X_t)+ \int_0^t  u(s,c_s) ds $$ is a supermartingale, and a martingale for the optimal policy.

We should thus expect a similar equation here, with the caveat that nature (or our cautiousness) chooses the worst parameters for drift $\mu$ and volatility $\sigma$. For given portfolio-consumption policy $(\pi,c)$, nature will thus minimize our utility, which leads locally to the \emph{uncertain HJB equation}

\begin{eqnarray}\label{UncertainHJB}
   &&\sup_{\pi, c} \inf_{(\mu,\sigma)\in \Theta} \Big\{u(t,c) - c v_x(t,x) + \pi x (\mu-r) v_x
   (t,x)\nonumber
   \\ &&\qquad \qquad\qquad + \frac{1}{2} \pi^2 x^2 \sigma^2 v_{xx} (t,x)\Big\} = 0 .
\end{eqnarray}

In fact, if we can find a suitable smooth function that solves this
adjusted HJB equation, we have solved our problem. The following
verification theorem states this fact in more detail.

\begin{theorem}\label{t1}
 Let $\varphi\in C^{1,2}((0,T)\times
\mathbb{R}^{+})$ be a solution of the following equation
\begin{eqnarray}\label{e3}
      &\sup\limits_{(\pi, c)\in \mathcal {B}\times\mathcal
{A}}\Big\{u(t,c)+\varphi_{t}(t,x)+xr\varphi_{x}(t,x)(1-\pi^{T}\mathbbold{1})-c\varphi_{x}(t,x)]\nonumber
\\&\quad +\inf\limits_{(q,Q)\in \Theta}\{\varphi_{x}(t,x)x\langle\pi,q\rangle
+\dfrac{1}{2}x^{2}\varphi_{xx}(t,x)\langle
A^{T}\pi\pi^{T}A,Q\rangle\}\Big\}=0,
\end{eqnarray}
with boundary condition
\begin{eqnarray*}
\varphi(T,x)= \Phi(T,x).
\end{eqnarray*}
Then we have
\begin{eqnarray*}
V(x_{0})&=&\varphi(0,x_0)= \sup\limits_{(\pi, c)\in \Pi\times C}
U(c, X).
\end{eqnarray*}

\end{theorem}

The above theorem delivers  the uncertain HJB equation for optimal consumption and optimal choice with very general specifications of
drift and volatility
   ambiguity. In our canonical case, these uncertainties are easily separated from each other (and are, in this sense, "independent''),
   but there are many more interesting possible specifications for a dependence between drift and volatility uncertainty.   For example,
   Epstein and
Ji (2013) consider
   \begin{eqnarray*}
\Theta=\Big\{ (\mu,\sigma^{2}): \mu=\mu_{min}+z,
\sigma^{2}=\sigma^{2}_{min}+\alpha z, 0\leq z\leq
\overline{z}\Big\},
\end{eqnarray*}
 where $\mu_{min}, \sigma^{2}_{min}, \alpha>0$ and
 $\overline{z}$ are fixed and deterministic
parameters.

\subsection{The Worst--Case Measure in the Canonical Model}

In this section, we will completely solve the ambiguity--averse investor's choice problem by reducing it to a suitable classical expected utility maximizer's problem. To do so, we will analyze the HJB equation in order to guess the worst--case prior. We will then verify that the value function under the worst--case prior solves our uncertain HJB equation as well.

In the canonical model, nature's minimization problem can be
explicitly solved. Let us have a look at the uncertain HJB equation
again. First of all, as in the classical case, it is natural to
expect that indirect utility is increasing, or $\varphi_x >0$,  and
(differentiably strictly) concave, or $\varphi_{xx}<0$.

In the canonical model, for the drift, we obtain then simply
 \begin{eqnarray*}
&&\inf\limits_{\mu\in [\underline{\mu},\overline{\mu}]}
\Big\{\varphi_{x}(t,x)x\langle\pi,\mu\rangle\Big\}\\ &=&
\varphi_{x}(t,x)x\sum\limits_{i=1}^{d}\pi^{i}\underline{\mu}^{i}1_{\{\pi^{i}>
0\}}+\varphi_{x}(t,x)x\sum\limits_{i=1}^{d}\pi^{i}\overline{\mu}^{i}1_{\{\pi^{i}\leq
0\}}.
\end{eqnarray*}
Clearly, nature decides for a low drift if we are long, and for a
high drift if we are short.

For the volatilities, as the value function is concave, we end up maximizing the potential volatility of returns.
For volatility, we immediately conclude that the nature always chooses its maximal possible value. Risk-- and ambiguity--averse investors use a cautious estimate for volatility, or add an ambiguity premium to their estimate.

Having solved nature's choice, we are left with a simple
maximization problem for consumption and portfolio weights, yet with
a kink in the linear part  at zero, when we change from short to
long position. We thus need to maximize
\begin{eqnarray*}
      &&u(t,c)+\varphi_{t}(t,x)+\varphi_{x}(t,x)xr-\varphi_{x}(t,x)c
+\varphi_{x}(t,x)x\sum\limits_{i=1}^{d}\pi^{i}(\underline{\mu}^{i}-r)1_{\{\pi^{i}>
0\}}\\&&\quad\quad\quad+\varphi_{x}(t,x)x\sum\limits_{i=1}^{d}\pi^{i}(\overline{\mu}^{i}-r)1_{\{\pi^{i}\leq
0\}}+\dfrac{1}{2}x^{2}\varphi_{xx}(t,x)\sum\limits_{i=1}^{d}(\pi^{i})^{2}(\overline{\sigma}^{i})^{2}
\end{eqnarray*}
over $\pi$ and $c$.

The solution for the portfolio depends on the relation of the riskless interest rate with
respect to the bounds for the drift.
 Let us write $$a(t,x)=- \frac{x\varphi_{xx}(t,x) }{\varphi_x(t,x)}$$ for the agent's indirect relative
 risk aversion.  The optimal portfolio choice anticipates the worst case scenarios. The investor
 evaluates the best long position and the best short position, and then make a choice of the better one.
 The optimal portfolio is
$$ \hat{\pi}^{i}=\frac{\underline{\mu}^{i}-r}{a(t,x) (\overline{\sigma}^i)^{2} }, i=1,\cdots, d,$$
if the lowest possible drift is above the riskless rate,
$\underline{\mu}^{i}>r$, this is a long position. And the optimal
portfolio choice is
$$ \hat{\pi}^{i}=\frac{\overline{\mu}^{i}-r}{a(t,x) (\overline{\sigma}^{i})^{2}} $$
if in contrast $\overline{\mu}^{i}<r$, and this is a short position.
The important case is the middle one, when ambiguity allows for
lower or higher drift than the interest rate; in this case, the
optimal portfolio does not invest into the uncertain asset,
$$\hat{\pi}^i=0\,.$$ This is because the long position is evaluated
by the lowest premium, and the short position is evaluated by the
highest premium. The return of wealth is strictly lower than the
riskless rate, which results in the nonparticipation of the risky
asset market.

When we compare the formulas for optimal portfolios to the classic Merton solution, we come  to the following conjecture: the investor
 behaves as if the lowest possible drift $\underline{\mu}^{i}$ was the real one if $\underline{\mu}^{i}>r$. If the interest rate belongs to the
 interval of possible drifts, then he behaves as if the drift was equal to the riskless rate (in this case, a standard risk-averse
 expected utility maximizer does not invest in the risky asset). Let us thus define the \emph{worst case parameters} as follows:
the worst case volatility is the highest possible volatility,
$\sigma^*=[\overline{\sigma}^{1},\cdots, \overline{\sigma}^{d}]$.
The worst drift depends on
 the relation of the interest rate $r$ with respect to $[\underline{\mu},\overline{\mu}]$:
\begin{equation*}
  \hat{\mu}^i = \left\{\begin{array}{cc}
  \underline{\mu}^{i},  & \mbox{if $\underline{\mu}^{i}>r $}; \\
r, &\ \qquad \mbox{if $\underline{\mu}^{i}\le r \le \overline{\mu}^{i} $};\\
\overline{\mu}^{i},  &\mbox{else}. \\
\end{array}\right.
\end{equation*}
We let $\mu^*=[\hat{\mu}^{1},\cdots, \hat{\mu}^{d}]$.  Let  the the
probability measure $P^*=P^{\mu^*,\sigma^*}$ the \emph{worst case
prior}.  $\varphi(0,x_{0})$ is the value function of an expected
utility maximizer using the worst-case prior.

\begin{theorem}\label{t2}
The ambiguity--averse investor chooses the same optimal policy as an
expected utility-maximizer with worst case prior $P^*$. In
particular, the value function $\varphi$ of an expected utility
maximizer with prior $P^*$ solves the uncertain HJB equation
(\ref{UncertainHJB}),
\begin{eqnarray*}
V(x_{0})&=&\varphi(0,x_0)=\sup\limits_{(\pi, c)\in \Pi\times C}
U(c,X),
\end{eqnarray*}
the optimal consumption rule is
\begin{eqnarray*}\label{}
    \hat{c}=v(\varphi_{x}(t,x)),
\end{eqnarray*}
where $v$ is the inverse of $u_{c}$, and
\begin{enumerate} {}
\item[(i)] if $r\leq\inf\limits_{i}\underline{\mu}^{i}$, then the
optimal portfolio choice is  $\hat{\mu}^{i}=\underline{\mu}^{i},
i=1,\cdots,d,$ and
\begin{eqnarray*}\label{}
  \hat{\pi}^{i}=-\frac{\varphi_{x}(t,x)}{\varphi_{xx}(t,x)x} \frac{\underline{\mu}^{i}-r}{(\overline{\sigma}^{i})^{2}}.
\end{eqnarray*}

\item[(ii)] if $\sup\limits_{i}\overline{\mu}^{i}\leq r$, then the
optimal portfolio choice is $\hat{\mu}^{i}=\overline{\mu}^{i},
i=1,\cdots,d,$ and
\begin{eqnarray*}\label{}
  \hat{\pi}^{i}=-\frac{\varphi_{x}(t,x)}{\varphi_{xx}(t,x)x} \frac{\overline{\mu}^{i}-r}{(\overline{\sigma}^{i})^{2}}.
\end{eqnarray*}

\item[(iii)]
if $\inf\limits_{i}\underline{\mu}^{i}<
r<\sup\limits_{i}\overline{\mu}^{i}$, then the optimal portfolio
choice is
 \begin{eqnarray*}\label{}
  \hat{\pi}^{i}=-\frac{\varphi_{x}(t,x)}{\varphi_{xx}(t,x)x}
  \frac{\underline{\mu}^{i}-r}{(\overline{\sigma}^{i})^{2}}1_{\{r\leq\underline{\mu}^{i} \}}
  -\frac{\varphi_{x}(t,x)}{\varphi_{xx}(t,x)x} \frac{\overline{\mu}^{i}-r}
  {(\overline{\sigma}^{i})^{2}}1_{\{\overline{\mu}^{i}\leq
r\}}.
\end{eqnarray*}

\end{enumerate}

\end{theorem}

The previous theorem allows to draw several interesting conclusions. First of all, as we have identified a worst--case measure, we have proved a minmax theorem.

\begin{corollary}
  We have the following minmax theorem:
  Let $\phi(P,x)$ denote the value function of an expected utility maximizer with belief $P$ and initial capital $x$. Let $v(x)$ be the ambiguity--averse investor's indirect utility function.
  Then
  $$v(x)=\min_{P\in\mathcal{P}} \phi(P,x)$$
  or
  $$\max_{(\pi,c) } \min_{P \in \mathcal{P}} E^P [\int_{0}^{T} u(s,
c_{s})ds+\Phi(T,X_T)] =
  \min_{P \in \mathcal{P}}     \max_{(\pi,c) } E^P [\int_{0}^{T} u(s,
c_{s})ds+\Phi(T,X_T)] \,.$$
\end{corollary}

The fact that the ambiguity--averse investor behaves as an expected utility maximizer under the worst--case measure $P^*$ does not imply that their demand functions are indistinguishable. Note that the worst--case measure is frequently the one where the drift is equal to the interest rate. Under such a belief, the expected utility maximizer does not invest at all into the risky asset, a result that is by now well established.

\subsection{Explicit solution for CRRA Utility}
In this subsection, we give   give explicit solutions for  Constant
Relative Risk Aversion (CRRA) Utility, i.e.,
$$u(t,c)=\dfrac{c^{1-\alpha}}{1-\alpha},
\Phi(T,x)=\dfrac{Kx^{1-\alpha}}{1-\alpha}, \alpha\neq 1.$$
 We have the following results. For the proof, see the appendix.
\begin{proposition}\label{ep1}

 \begin{enumerate} {}
\item[(i)] If $r\leq\inf\limits_{i}\underline{\mu}^{i}$, then the optimal
 consumption and portfolio rules are given by the
following
\begin{eqnarray*}\label{}
  \hat{c}=\Big[K^{\alpha^{-1}}e^{\beta\alpha^{-1}(T-t)}+\alpha
\beta^{-1}(e^{\beta\alpha^{-1}(T-t)}-1)\Big]^{-1}x,
\end{eqnarray*}
 and
\begin{eqnarray*}\label{}
  \hat{\pi}^{i}=\frac{1}{\alpha
  }\frac{\underline{\mu}^{i}-r}{(\overline{\sigma}^{i})^{2}},\
  i=1,\cdots, d,
\end{eqnarray*}
where
\begin{eqnarray*}\label{}
  \beta=\big[r+\sum\limits_{i=1}^{d}\frac{(\underline{\mu}^{i}-r)^{2}}{2\alpha(\overline{\sigma}^{i})^{2}}\big](1-\alpha).
\end{eqnarray*}

\item[(ii)] if $\sup\limits_{i}\overline{\mu}^{i}\leq r$,
then the optimal consumption and portfolio rules are given by the
following
\begin{eqnarray*}\label{}
  \hat{c}=\Big[K^{\alpha^{-1}}e^{\beta\alpha^{-1}(T-t)}+\alpha
\beta^{-1}(e^{\beta\alpha^{-1}(T-t)}-1)\Big]^{-1}x,
\end{eqnarray*}
 and
\begin{eqnarray*}\label{}
  \hat{\pi}^{i}=\frac{1}{\alpha
  }\frac{\overline{\mu}^{i}-r}{(\overline{\sigma}^{i})^{2}},\
  i=1,\cdots, d,
\end{eqnarray*}
where
\begin{eqnarray*}\label{}
  \beta=\big[r+\sum\limits_{i=1}^{d}\frac{(\overline{\mu}^{i}-r)^{2}}{2\alpha(\overline{\sigma}^{i})^{2}}\big](1-\alpha).
\end{eqnarray*}

\item[(iii)]
if $\inf\limits_{i}\underline{\mu}^{i}<
r<\sup\limits_{i}\overline{\mu}^{i}$, then the  optimal consumption
and portfolio rules are given by the following
\begin{eqnarray*}\label{}
  \hat{c}=\Big[K^{\alpha^{-1}}e^{\beta\alpha^{-1}(T-t)}+\alpha
\beta^{-1}(e^{\beta\alpha^{-1}(T-t)}-1)\Big]^{-1}x,
\end{eqnarray*}
 and
 \begin{eqnarray*}\label{}
  \hat{\pi}^{i}=\frac{1}{\alpha}
  \frac{\underline{\mu}^{i}-r}{(\overline{\sigma}^{i})^{2}}1_{\{r\leq\underline{\mu}^{i} \}}
  +\frac{1}{\alpha} \frac{\overline{\mu}^{i}-r}
  {(\overline{\sigma}^{i})^{2}}1_{\{\overline{\mu}^{i}\leq
r\}},
\end{eqnarray*}
where
\begin{eqnarray*}\label{}
  \beta=\big[r+\sum\limits_{i=1}^{d}\frac{(\underline{\mu}^{i}-r)^{2}}{2\alpha(\overline{\sigma}^{i})^{2}}\big](1-\alpha)1_{\{r\leq\underline{\mu}^{i} \}}+
  \big[r+\sum\limits_{i=1}^{d}\frac{(\overline{\mu}^{i}-r)^{2}}{2\alpha(\overline{\sigma}^{i})^{2}}\big](1-\alpha)
  1_{\{\overline{\mu}^{i}\leq
r\}}.
\end{eqnarray*}
 In particular, if $\underline{\mu}^{i}<
r<\overline{\mu}^{i}$, then the optimal portfolio choice is
\begin{eqnarray*}\label{}
  \hat{\pi}^{i}=0.
\end{eqnarray*}

\end{enumerate}
\end{proposition}

\subsection{Comparative Statics}

From the above results in the above subsection we obtain immediately 
following comparative statics.

\begin{proposition}
  Let drift ambiguity be given by $[\mu_0-\kappa, \mu_0 + \kappa]$ for some $\kappa>0$. As $\kappa$ increases,
  asset holdings decrease. After some critical level of ambiguity $\kappa^*$,
  refrains from trading the asset altogether.
\end{proposition}

We have here the well--known phenomenon that high uncertainty about mean returns keeps ambiguity--averse investors away form the asset market, as \cite{DowWerlang92} first pointed out. 

\begin{proposition}
  The exposure of investors decreases with ambiguity: for parameter sets $\Theta \subset \hat{\Theta}$,
  let $\pi$ and $\hat{\pi}$ denote the optimal portfolio choices. Then $\| \pi \| \geq \| \hat{\pi} \|$.
\end{proposition}

This result follows from the fact that the investor always works with the maximal volatility. If the asset is profitable, he uses the minimal mean excess return, and if she is going short, she uses the maximal mean return in computing the portfolio. The absolute amount of assets held optimally thus decreases with ambiguity. 

\section{Interest rate uncertainty}\label{section4}

A fixed and known interest rate, or in other words, a flat term
structure, is not a reasonable assumption for long-term investors.
Investors face considerable uncertainty about the short rate; on the
one hand, stochastic demand for credit leads to short term
variability, on the other hand, the short rate is partially
determined by central bank policies. The latter are known to be
quite ambiguous, and sometimes deliberately so as central bankers
have strong incentives to conceal their real objectives. We thus
also allow for interest rate uncertainty.

Introducing Knightain uncertainty about the short rate
\emph{requires} the use of singular measures: if we model the
possibility that the bond dynamics satisfy
$$dP_t= r_t P_t dt$$
under one measure, and
$$dP_t = \hat{r}_t P_t dt $$
under another, for a different short rate $\hat{r}$, then these
measures need to be singular to each other. This is in contrast to
the well--studied models of drift ambiguity for the uncertain asset
where the presence of the noise term allows to work with equivalent
probability measures.

But we are already used to work with singular measures, and so  our
framework can be extended to cover such Knightian uncertainty about
the short rate as well. We model the ambiguity about the short rate
via an interval $[\underline{r},\overline{r}]$, similar to the
ambiguity about drift and volatility.

Interest rate uncertainty leads to some new phenomena. The most
interesting case arises, in our view, when the uncertain asset is
profitable, but interest rate uncertainty is high. Then it is
optimal not to participate in the market for bonds at all and to put
all capital into the uncertain asset. We thus obtain
non--participation in the credit market; a phenomenon that we have
seen during the financial crisis as well. Of course, we do not model
or explain the origin of such uncertainty here, but we show that
interest rate uncertainty can play an important role in asset
decisions.

Let us now come to the formal model. In a first step,  as in Section
2, we construct a set of priors. For  $\theta=(\mu,\sigma) $ and
$r$, which are an $\mathcal {F}$-progressively measurable processes
with values in
$\Theta=[\underline{\mu},\overline{\mu}]\times[\underline{\sigma}^{2},\overline{\sigma}^{2}]$
and $[\underline{r},\overline{r}]$, respectively, we consider
stochastic differential equation
$$dX^\theta_t = \mu_t dt + \sigma_t dB_t, X_0=0,$$
and  $$dY^r_t = r_t dt,\ Y_0=0,$$  under our reference measure
$P_0$.

 We let $P^{r,\theta}$ be the
distribution of $(X^\theta, Y^r)$, i.e.
$$P^{r,\theta}(A)=P_0((X^\theta, Y^r) \in A), $$ for all $A \in \mathcal {F}_{T}$.

Let $\mathcal {P}_{0}$ be consist  of all probability measures
$P^\theta$ constructed in this way. Our set of priors $\mathcal {P}$
is the closure of $\mathcal {P}_{0}$ under the topology of weak
convergence. The set of priors leads naturally to a sublinear
expectation:
$$\mathbb{\hat{E}}[\cdot]=\sup\limits_{P\in\mathcal {P}} E_P[\cdot].$$
 The
sublinear function  $G: \mathbb{R}^{3}\times \rightarrow\mathbb{R}:$
\begin{eqnarray*}
G(p_{1},p_{2},p_{3})=\sup\limits_{(r,\mu,\sigma)\in
[\underline{r},\overline{r}]\times[\underline{\mu},\overline{\mu}]\times[\underline{\sigma}^{2},\overline{\sigma}^{2}]
}\Big\{ rp_{1}+\mu p_{2}+\dfrac{1}{2}\sigma^{2}p_{3}\Big\}.
\end{eqnarray*}

 As introduced in
the Section 2,  let  $(B,b,\hat{b})$ be a pair of random vectors
under $\mathbb{\hat{E}}$ such that $B$ is a $G$-Brownian motion and
 $(b,\hat{b})$ is a $G$-distributed process, which  has mean uncertainty. $B$ is a $G$-normal distributed process,
 which just has volatility uncertainty.

  In a financial market, we consider the optimal consumption and
portfolio choice, not only with ambiguity about returns and
volatility, but also with  interest rate uncertainty. The price of
the riskless asset is  now  defined by
\begin{eqnarray*}
dP_{t}=P_{t}d\hat{b}_{t},
\end{eqnarray*}
where $\hat{b}$ is a $G$-distributed process, and $\hat{b}_{t}$ has
 mean uncertainty  $[\underline{r}t,\overline{r}t]$.

In this section, we just consider one risky asset in the financial
market. The classical ``risky'' assets where we assume that expected
return and volatility are unknown. The uncertain asset prices evolve
as
$$dS_t = S_t dR_t $$ and we model the return dynamics via
\begin{eqnarray*}
dR_{t}=db_{t} +dB_{t}.
\end{eqnarray*}
As we did in Section 2, we
 can define the consumption and trading opportunities, and utility
 in the same way.

We consider the case corresponding to the Constant Relative Risk
Aversion (CRRA) Utility, i.e.,
$$u(t,c)=\dfrac{c^{1-\alpha}}{1-\alpha},
\Phi(T,x)=\dfrac{Kx^{1-\alpha}}{1-\alpha}, \alpha>0, \alpha\neq 1.$$

The following three potentially optimal portfolio shares play a role
as candidate optimal policies in the following. First,
$$\pi^1 = \frac{\overline{\mu}-\underline{r}}{\alpha \overline{\sigma}^2}$$
corresponds to the case when maximal drift and maximal interest rate
are the worst case parameters. Similarly, we define
$$\pi^2 = \frac{\underline{\mu}-\underline{r}}{\alpha \overline{\sigma}^2}$$
and
$$\pi^3 = \frac{\underline{\mu}-\overline{r}}{\alpha \overline{\sigma}^2}\,.$$
Note that $$ \pi^1 \ge \pi^2 \ge \pi^3.$$

We have the following main result in this section.

\begin{theorem}\label{5t}
The optimal consumption--investment policies under interest rate
uncertainty can be divided into five cases:
\begin{enumerate}
\item If $\pi^1 \ge 0$,
\begin{enumerate}
\item and $\pi^2 \le 0$,
non--participation in the asset market, or $\pi^*=0$, is optimal,
\item and $0<\pi^2 < 1$, the investor goes long and saves, i.e. $\pi^*=\pi^2$,
\item and for $\pi^2 \ge 1$, we have
\begin{enumerate}
\item in case $\pi^3 < 1$,  the investor puts all capital in the uncertain asset and does not
participate in the credit market, or $\pi^*=1$,
\item in case $\pi^3 \ge 1$, the  investor goes long and borrows (leveraged consumption) and $\pi^*=\pi^3$.
\end{enumerate}
\end{enumerate}
\item if $\pi^1 < 0$, the  investor goes short and saves (leveraged consumption) and $\pi^*=\pi^1$.
\end{enumerate}
\end{theorem}

For the above theorem, see the following figures.

 If
$\underline{\mu}\leq\underline{r}$, then
\[\underline{\mu} \frac{\ \ \pi^*=0\ \
}{}\ \overline{\mu}\ \frac{\ \
\pi^*=\frac{\underline{\mu}-\overline{r}}{\alpha
\overline{\sigma}^2}\ \ }{}\hskip-1.6mm\xrightarrow[\
\underline{r}]{}.
\]

If \noindent $\underline{\mu}\geq\underline{r}$, then

\[\underline{r}-\alpha
\overline{\sigma}^2\ \frac{\ \
\pi^*=\frac{\underline{\mu}-\underline{r}}{\alpha
\overline{\sigma}^2}\ \ }{}\ \underline{r}\ \frac{\ \ \pi^*=1\ \
}{}\ \overline{r}\ \frac{\ \
\pi^*=\frac{\underline{\mu}-\overline{r}}{\alpha
\overline{\sigma}^2}\ \ }{}\hskip-2.2mm\xrightarrow[\
\underline{\mu}-\alpha \overline{\sigma}^2]{}.
\]

The details of the proof are in the appendix.  Our model allows to
quantify under what conditions  non-participation in the credit
market is optimal for ambiguity-averse investors. This occurs when
the highest possible drift exceeds the highest possible interest
rate, and investment in the uncertain asset is thus potentially
profitable, but when the lowest possible drift, adjusted by a
mean-variance term involving risk aversion, $\underline{\mu}-\alpha
\overline{\sigma}^2 $, belongs to the interval of possible interest
rates $[\underline{r}, \overline{r}]$.

\vskip15mm

\noindent{\Large \textbf{Appendix}}
\appendix
\section{$G$-Brownian motion}

Peng (2007) introduced the theory of $G$-Brownian motion. For the
convenience of the readers, we recall the basic definitions and some
results of the theory of $G$-Brownian motion.

Let $\Omega$ be a given nonempty set and $\mathcal {H}$ be a linear
space of real functions defined on $\Omega$ such that if
$x_{1},\cdot\cdot\cdot,x_{n}\in \mathcal {H}$, then
$\varphi(x_{1},\cdot\cdot\cdot,x_{n})\in \mathcal {H}$, for each
$\varphi\in C_{l,lip}(\mathbb{R}^{n})$. Here
$C_{l,lip}(\mathbb{R}^{n})$ denotes the linear space of functions
$\varphi$ satisfying
$$|\varphi(x)-\varphi(y)|\leq C(1+|x|^{n}+|y|^{n})|x-y|,\ \text{for all}\ x, y\in \mathbb{R}^{n},$$
for some $C>0$ and $\ n\in\mathbb{N}$, both depending on $\varphi$.
The space $\mathcal {H}$ is considered as a set of random variables.

\begin{definition}
{\bf A sublinear expectation} $\mathbb{\hat{E}}$ on $\mathcal {H}$
is a functional $\mathbb{\hat{E}}:\mathcal {H}\mapsto \mathbb{R}$
satisfying the following properties: for all $X, Y \in \mathcal
{H}$, we have
\begin{enumerate}
\item[(i)] {\bf Monotonicity:} If $X\geq Y $, then
$\mathbb{\hat{E}}[X]\geq\mathbb{\hat{E}}[Y]$.
\item[(ii)] {\bf Preservation of constants:}
$\mathbb{\hat{E}}[c]=c$, for all $c\in \mathbb{R}$.
\item[(iii)] {\bf Subadditivity:}
$\mathbb{\hat{E}}[X]-\mathbb{\hat{E}}[Y]\leq\mathbb{\hat{E}}[X-Y]$.
\item[(iv)] {\bf Positive homogeneity:}
$\mathbb{\hat{E}}[\lambda X]=\lambda\mathbb{\hat{E}}[X],$ for all $
\lambda \geq 0$.
\end{enumerate}
The triple $(\Omega, \mathcal {H}, \mathbb{\hat{E}})$ is called a
sublinear expectation space.
\end{definition}

\begin{remark}
    The sublinear expectation space can be regarded as a generalization of the classical
probability space $(\Omega, \mathcal {F}, \mathbb{P})$ endowed with
the linear expectation associated with $\mathbb{P}$.
\end{remark}

\begin{definition} In a sublinear expectation space $(\Omega, \mathcal {H},
\mathbb{\hat{E}})$, a random vector $Y=(Y_{1},\cdots, Y_{n}),
Y_{i}\in \mathcal {H}$, is said to be independent under $
\mathbb{\hat{E}}$ of another random vector $X=(X_{1},\cdots, X_{m}),
X_{i}\in \mathcal {H}$, denoted by $X\perp Y,$  if for each test
function $\varphi \in C_{l,lip}(\mathbb{R}^{m+n})$ we have
$$\mathbb{\hat{E}}[\varphi(X, Y)]=\mathbb{\hat{E}}[\mathbb{\hat{E}}[\varphi(x, Y)]_{x=X}].$$
\end{definition}

\begin{definition} In a sublinear expectation space $(\Omega, \mathcal {H},
\mathbb{\hat{E}})$,  $X$ and $Y$ are called identically distributed,
and denoted by $X \stackrel{d}{=} Y$,   if for each  $\varphi \in
C_{l,lip}(\mathbb{R}^{n})$ we have
$$\mathbb{\hat{E}}[\varphi(X)]=\mathbb{\hat{E}}[\varphi(Y)].$$
\end{definition}

\begin{definition}{\bf($G$-distribution)}
   A pair of random variables $(X,\eta)$ in a sublinear expectation space $(\Omega, \mathcal {H},
\mathbb{\hat{E}})$ is called $G$-distributed, if for all $a, b\geq
0, $
$$(aX+b\bar{X}, a^{2}\eta+b^{2}\bar{\eta})\stackrel{d}{=}\sqrt{a^{2}X+b^{2}}\bar{X}+(a^{2}+b^{2})\eta,$$
where $(\bar{X},\bar{\eta})$ is an independent copy of $(X,\eta)$,
i.e., $(\bar{X},\bar{\eta})\stackrel{d}{=}(X,\eta)$ and
$(\bar{X},\bar{\eta})\perp (X,\eta)$.
\end{definition}

 If $(X,\eta)$ is $d$ dimensional  $G$-distributed, for $\varphi \in
C_{l,lip}(\mathbb{R}^{d})$, let us define
$$u(t, x,y):=\mathbb{\hat{E}}[\varphi(x+\sqrt{t}\xi,y+t\eta)], \quad (t,x,y)\in [0, \infty)
\times\mathbb{R}^{d}\times\mathbb{R}^{d},$$ is the   solution  of
the following parabolic partial differential equation:
$$\left\{\begin{array}{l l}
      \partial_{t}u(t, x)=G(D_{y}u(t, x,y),D_{xx}^{2}u(t, x,y)),\quad (t, x)\in[0,
\infty)\times\mathbb{R}^{d},\\
       u(0, x)=\varphi(x).
         \end{array}
  \right.$$
 Here  $G$  is the following sublinear function:
 $$G(A)=\mathbb{\hat{E}}[\dfrac{1}{2}\langle AX,X\rangle+\langle p,\eta\rangle], \quad (p,A) \in
\mathbb{R}^{d}\times \mathbb{S}_{d},$$
 where $\mathbb{S}_{d}$ is the collection of $d\times d$ symmetric
 matrices. There exists a bounded and
closed subset $\Theta $ of $\mathbb{R}^{d}\times\mathbb{R}^{d\times
d}$ such that
\begin{eqnarray*}
G(p,A)=\sup\limits_{(q,Q)\in\Theta}\Big\{\langle
p,q\rangle+\dfrac{1}{2}tr(AB)\Big\}, \ \text{for }\
(p,A)\in\mathbb{R}^{d}\times \mathbb{S}_{d}.
\end{eqnarray*}

Let $\Omega=C^{2d}_{0}(\mathbb{R^{+}})$ be the space of all
$\mathbb{R}^{d}$-valued continuous paths $(\omega_{t})_{t\in
\mathbb{R^{+}}}$ with $\omega_{0}=0$, equipped with the distance
$$\rho(\omega^{1}, \omega^{2})
=\sum\limits_{i=1}^{\infty}2^{-i}\Big[(\max\limits_{t\in[0,i]}|\omega_{t}^{1}-\omega_{t}^{2}|)\wedge1\Big],
\ \omega^{1}, \omega^{2}\in\Omega.$$

For each $t\in [0,+\infty),$ we set
$\omega_{t}=\{\omega_{\cdot\wedge t}, \omega\in \Omega\}.$ We will
consider the canonical process
$\hat{B}_{t}(\omega)=(B_{t},b_{t})(\omega)=\omega_{t}, t\in
[0,+\infty), \omega\in\Omega.$

 For each $T > 0$, we consider the following space of
random variables:
\begin{eqnarray*}
L_{ip}(\Omega_{T}):=\Big\{\varphi(\omega_{{t_{1}}}
\cdots, \omega_{{t_{m}}}) \ | \ t_{1}, \cdots, t_{m}\in [0,T], \\
\ \varphi\in C_{l,lip}(\mathbb{R}^{d\times m}), \ m\geq1\Big\}.
\end{eqnarray*}
Obviously, it holds that $L_{ip}(\Omega_{t}) \subseteq
L_{ip}(\Omega_{T})$, for all $t\leq T<\infty$.   We further define
$$L_{ip}(\Omega)=\bigcup_{n=1}^{\infty}L_{ip}(\Omega_{n}).$$

 For each $X\in L_{ip}(\Omega)$ with
$$X=\varphi(\hat{B}_{t_{1}}-\hat{B}_{t_{0}}, \hat{B}_{t_{2}}-\hat{B}_{t_{1}}, \cdots,
 \hat{B}_{t_{m}}-\hat{B}_{t_{m-1}})$$
for some $m\geq 1, \varphi\in C_{l,lip}(\mathbb{R}^{2d\times m})$
and $0=t_{0}\leq t_{1}\leq\cdots \leq t_{m}<\infty$, we set
\begin{eqnarray*}
&&\mathbb{\hat{E}}[\varphi(\hat{B}_{t_{1}}-\hat{B}_{t_{0}},
\hat{B}_{t_{2}}-\hat{B}_{t_{1}}, \cdots, \hat{B}_{t_{m}}-\hat{B}_{t_{m-1}})]\\
&&=\mathbb{\widetilde{E}}[\varphi(\sqrt{t_{1}-t_{0}}\xi_{1},
(t_{1}-t_{0})\eta_{1} , \cdots, \sqrt{t_{m}-t_{m-1}}\xi_{m},
(t_{m}-t_{m-1})\eta_{m})],
\end{eqnarray*}
where $\{(\xi_{1},\eta_{1}),\cdots, (\xi_{m},\eta_{m})\}$ is a
random vector in a sublinear expectation space $(\widetilde{\Omega},
\widetilde{\mathcal {H}}, \widetilde{\mathbb{E}})$ such that
$(\xi_{i},\eta_{i})$ is $G$-distributed $(\xi_{i+1},\eta_{i+1})$ is
independent of $\{(\xi_{1},\eta_{1}),\cdots, (\xi_{i},\eta_{i})\}$,
for every $i=1, 2, \cdots, m-1$.

The related conditional expectation of
$X=\varphi(\hat{B}_{t_{1}}-\hat{B}_{t_{0}},
\hat{B}_{t_{2}}-\hat{B}_{t_{1}}, \cdots,
\hat{B}_{t_{m}}-\hat{B}_{t_{m-1}})$ under $\Omega_{t_{j}}$ is
defined by
\begin{eqnarray*}
\mathbb{\hat{E}}[X|\Omega_{t_{j}}]&=&\mathbb{\hat{E}}[\varphi(\hat{B}_{t_{1}}-\hat{B}_{t_{0}},
\hat{B}_{t_{2}}-\hat{B}_{t_{1}}, \cdots, \hat{B}_{t_{m}}-\hat{B}_{t_{m-1}})|\Omega_{t_{j}}]\\
&=&\psi(\hat{B}_{t_{1}}-\hat{B}_{t_{0}},
\hat{B}_{t_{2}}-\hat{B}_{t_{1}}, \cdots, B_{t_{j}}-B_{t_{j-1}}),
\end{eqnarray*}
where
\begin{eqnarray*}
&&\psi(x_{1}, x_{2}, \cdots,
x_{j})\\&=&\mathbb{\widetilde{E}}[\varphi(x_{1}, x_{2}, \cdots,
x_{j}, \sqrt{t_{j+1}-t_{j}}\xi_{j+1}, (t_{j+1}-t_{j})\eta_{j+1}
\cdots, \\&& \qquad \sqrt{t_{m}-t_{m-1}}\xi_{m},
(t_{m}-t_{m-1})\eta_{m})],
\end{eqnarray*}
for $(x_{1}, x_{2}, \cdots, x_{j})\in\mathbb{R}^{j}, 0\leq j\leq m.$

\vspace{2mm} For $p\geq 1$,
$\|X\|_{p}=\mathbb{\hat{E}}^{\frac{1}{p}}[|X|^{p}]$,  $X \in
L_{ip}(\Omega),$ defines a norm on  $L_{ip}^{0}(\Omega)$. Let
$L^{p}(\Omega)$ (resp. $L^{p}(\Omega_{t})$) be the completion of
$L_{ip}(\Omega)$ (resp. $L_{ip}(\Omega_{t})$) under the norm
$\|\cdot\|_{p}$. Then the space $(L^{p}(\Omega), \|\cdot\|_{p})$ is
a Banach space and the operators $\mathbb{\hat{E}}[\cdot] $ (resp.
$\mathbb{\hat{E}}[\cdot|\Omega_{t}]$) can be continuously extended
to the Banach space $L^{p}(\Omega)$ (resp. $L^{p}(\Omega_{t})$).
Moreover, we have $L^{p}(\Omega_{t})\subseteq
L^{p}(\Omega_{T})\subset L^{p}(\Omega)$, for all $0\leq t \leq T
<\infty$.

\begin{definition}{\bf($G$-normal distribution)}
 Let  $\Gamma$ be a   given non-empty, bounded and closed subset of  $\mathbb{R}^{d\times d}$. A random
 vector $\xi$ in a sublinear expectation space $(\Omega, \mathcal {H},
\mathbb{\hat{E}})$ is said to be $G$-normal distributed, denoted by
$\xi\sim \mathcal {N}(0, \Gamma)$, if for each $\varphi \in
C_{l,lip}(\mathbb{R}^{d})$, the following function defined by
$$u(t, x):=\mathbb{\hat{E}}[\varphi(x+\sqrt{t}\xi)], \quad (t,x)\in [0, \infty)\times\mathbb{R}^{d},$$
is the unique  viscosity solution  of the following parabolic
partial differential equation:
\begin{eqnarray}\label{heat1}
\left\{\begin{array}{l l}
      \dfrac{\partial u}{\partial t}=G(D^{2}u),\quad (t, x)\in[0,
\infty)\times\mathbb{R}^{d},\\
       u(0, x)=\varphi(x).
         \end{array}
  \right.
\end{eqnarray}
 Here  $D^{2}u$  is the Hessian matrix of $u$, i.e.,
 $D^{2}u=(\partial^{2}_{x^{i}x^{j}}u)_{i,j=1}^{d}$, and
 $$G(A)=\frac{1}{2}\sup\limits_{\gamma\in \Gamma}tr(\gamma\gamma^{T} A), \quad A \in \mathbb{S}_{d}.$$
\end{definition}

\begin{example}
In one dimensional case, i.e., $d=1$,  we take
$\Gamma=[{\underline{\sigma}^{2}, \overline{\sigma}}^{2}]$, where
$\underline{\sigma}$ and $\overline{\sigma}$ are constants  with
$0\leq\underline{\sigma}\leq \overline{\sigma}.$ Then equation
(\ref{heat1}) has the following form
$$\left\{\begin{array}{l l}
       \dfrac{\partial u}{\partial t}=\dfrac{1}{2}[\overline{\sigma}^{2}
      (\partial_{xx}^{2}u)^{+}-\underline{\sigma}^{2}(\partial_{xx}^{2}u)^{-}],\quad (t, x)\in[0,
\infty)\times\mathbb{R},\\
       u(0, x)=\varphi(x).
         \end{array}
  \right.$$
If $\underline{\sigma}= \overline{\sigma}$, the $G$-normal
distribution is the classical normal distribution.
\end{example}

\begin{example}
In multidimensional case,  we consider one typical case when
$$\Gamma=\Big\{diag[\gamma^{1},\cdots,\gamma^{d}],
\gamma^{i}\in[(\underline{\sigma}^{i})^{2},(\overline{\sigma}^{i})^{2}],
i=1,\cdots,d\Big\},$$  where $\underline{\sigma}^{i}$ and
$\overline{\sigma}^{i}$ are constants  with
$0\leq\underline{\sigma}^{i}\leq \overline{\sigma}^{i}.$ Then
equation (\ref{heat1}) has the following form
$$\left\{\begin{array}{l l}
      \dfrac{\partial u}{\partial t}=\dfrac{1}{2}\sum\limits_{i=1}^{d}[(\overline{\sigma}^{i})^{2}
      (\partial_{x^{i}x^{i}}^{2}u)^{+}-(\underline{\sigma}^{i})^{2}(\partial_{x^{i}x^{i}}^{2}u)^{-}], \\
       u(0, x)=\varphi(x).
         \end{array}
  \right.$$
\end{example}

\begin{definition}
 A process $B=\{B_{t},t\geq 0\}$ in a sublinear expectation space $(\Omega, \mathcal {H},
 \mathbb{\hat{E}})$ is called a $G$-Brownian motion, if the following
 properties are satisfied:
\begin{enumerate}
\item[(i)] $B_{0}=0$;
\item[(ii)] for each $t, s \geq0$, the difference $B_{t+s}-B_{t}$ is
$\mathcal {N}(0, \ \Gamma s)$-distributed and is independent of
$(B_{t_{1}}, \cdots, B_{t_{n}})$, for all $n\in\mathbb{N}$ and
$0\leq t_{1}\leq\cdots \leq t_{n}\leq t$.
\end{enumerate}
\end{definition}

\begin{definition}{\bf(Maximal distribution)}
 Let  $\Lambda$ be a   given non-empty, bounded and closed subset of  $\mathbb{R}^{d}$. A random
 vector $\xi$ in a sublinear expectation space $(\Omega, \mathcal {H},
\mathbb{\hat{E}})$ is said to be Maximal distributed, denoted by
$\xi\sim \mathcal {N}(\Lambda, \{0\})$, if for each $\varphi \in
C_{l,lip}(\mathbb{R}^{d})$, the following function defined by
$$u(t, x):=\mathbb{\hat{E}}[\varphi(x+ t\xi)], \quad (t,x)\in [0, \infty)\times\mathbb{R}^{d},$$
is the unique  viscosity solution  of the following parabolic
partial differential equation:
\begin{eqnarray*}\label{}
\left\{\begin{array}{l l}
      \dfrac{\partial u}{\partial t}=g(Du),\quad (t, x)\in[0,
\infty)\times\mathbb{R}^{d},\\
       u(0, x)=\varphi(x),
         \end{array}
  \right.
\end{eqnarray*}
where   $Du=(\partial_{x^{i}u})_{i=1}^{d}$, and
 $$g(p)=\frac{1}{2}\sup\limits_{q\in \Lambda} \langle p,q\rangle, \quad p \in \mathbb{R}^{d}.$$
\end{definition}

\begin{proposition}
If $b_{t}\sim \mathcal {N}([\underline{\mu}t, \overline{\mu}t],
\{0\})$, where $\underline{\mu}$ and $ \overline{\mu}$ are constants
with $\underline{\mu}\leq\overline{\mu}$, then for $\varphi \in
C_{l,lip}(\mathbb{R})$
\begin{eqnarray*}
 \mathbb{\hat{E}}[\varphi(b_{t})]=\sup\limits_{v\in [\underline{\mu}t,
 \overline{\mu}t]}\varphi(vt).
\end{eqnarray*}
\end{proposition}

\begin{proposition}
{\bf (It\^{o}'s formula)} Let $b_{t}\sim \mathcal
{N}([\underline{\mu}t, \overline{\mu}t], \{0\})$ and $B_{t}\sim
\mathcal {N}([\underline{\sigma}^{2}t, \overline{\sigma}^{2}t],
\{0\})$ where $\underline{\mu}$ and $ \overline{\mu}$ are constants
with $\underline{\mu}\leq\overline{\mu}$, and $\underline{\sigma}$
and $ \overline{\sigma}$ are constants with
$\underline{\sigma}\leq\overline{\sigma}$. Then for  $\varphi \in
C^{2}(\mathbb{R})$ and
$$X_{t}=X_{0}+\int_{0}^{t}\alpha
_{s}db_{s}+\int_{0}^{t}\beta_{s}dB_{s}, \ \text{for all} \
t\in[0,T],$$ where $\alpha$ in $M^{1}$ and $\beta \in M^{2}$, we
have
\begin{align*}
\varphi(X_{t})-\varphi(X_{0})   =& \int_{0}^{t}\partial_{x}\varphi(X_{u}%
)\beta_{u}dB_{u}+\int_{0}^{t}\partial_{x}\varphi(X_{u})\alpha_{u}db_{u}\\
& +\int_{0}^{t} \frac{1}{2}\partial
_{xx}^{2}\varphi(X_{u})\beta_{u}^{2} d\langle B\rangle_{u},\ \ 0\leq
t\leq T.
\end{align*}
\end{proposition}

\section{Construction of the Set of Priors}

In an ambiguous world, the investor is uncertain about the law that
governs the price dynamics of risky assets. We thus do not fix a
probability measure ex ante. We set up the canonical model in such a continuous--time Knightian setting first.

As we want to study the standard Samuelson--Merton consumption--portfolio problem in the Knightian case,  we look at asset prices with continuous sample paths.
We  let  $C([0,T])$
be the set of  all continuous paths with values in $\mathbb R^d$
over the finite time horizon $[0,T]$ endowed with the sup
norm. Our   state space is
\begin{eqnarray*}
\Omega_{0}=\Big\{\omega: \omega\in C([0,T]), \omega_{0}=0\Big\}.
\end{eqnarray*}
The coordinate process $B=(B_{t})_{t\geq 0}$ is
$B_{t}(\omega)=\omega_{t}$.

As in the classic case, the coordinate process
$B_t(\omega)=\omega(t)$ will play the role of noise, but here it
will be uncertain, rather than probabilistic noise; ambiguous, or,
as Peng calls it,  $G$--Brownian motion. In order to model such
ambiguous Brownian motion, we construct a set of priors. We take as
a starting point the classic Wiener measure $P_0$ under which $B$ is
a standard Brownian motion. Note that $P_0$ does not reflect the
investor's view of the world; it merely plays the role of a
construction tool for the set of priors.

Let  $\mathcal {F}=(\mathcal
{F}_{t})_{t\geq 0}$ denote the filtration generated by $B$, completed by all
$P_0$-null sets.

In the continuous--time diffusion framework, essentially two
parameter processes describe all uncertainty, drift and volatility.
We thus model ambiguity with the help of a convex and compact subset
$\Theta \subset \mathbb R^d \times \mathbb R^{d \times d}$. The
investor is not sure about the exact value or distribution of the
drift process $\mu=(\mu_t) $ with values in $\mathbb R^d$ nor about
the exact value or distribution of the volatility process
$\sigma=(\sigma_t)$ with values in $\mathbb R^{d \times d}$.

For every ``hypothesis'' $\theta=(\mu,\sigma) $, an $\mathcal
{F}$-progressively measurable process with values in $\Theta$, the
stochastic differential equation
$$dX_t = \mu_t dt + \sigma_t dB_t, X_0=0$$ has a unique solution $X^\theta$ under our
reference measure $P_0$. We let $P^\theta$ be the distribution  of
$X^\theta$, i.e.
$$P^\theta(A)=P_0(X^\theta \in A) $$ for all $A \in \mathcal {F}_{T}$.

Let $\mathcal {P}_{0}$ be consist  of all probability measures
$P^\theta$ constructed in this way. Our set of priors $\mathcal {P}$
is the closure of $\mathcal {P}_{0}$ under the topology of weak
convergence. Ambiguous volatility gives rise to nonequivalent
priors. For example, let $P^{\overline{\sigma}}$ and
$P^{\underline{\sigma}}$ be the distribution of the processes
$(\overline{\sigma}B_{t})_{t\geq 0}$ and
$(\underline{\sigma}B_{t})_{t\geq 0}$, respectively. Then
$P^{\overline{\sigma}}$ and $P^{\underline{\sigma}}$ are mutually
singular, i.e.,
\begin{eqnarray*}
P^{\overline{\sigma}}(\langle
B\rangle_{T}=\overline{\sigma}^{2}T)=P^{\underline{\sigma}}(\langle
B\rangle_{T}=\underline{\sigma}^{2}T)=1,
\end{eqnarray*}
where the quadratic variation process of $B$ is defined as follows,
for $0=t_{1}\leq\cdots<t_{m}=T$ and $\Delta t_{k}=t_{k+1}-t_{k},$
\begin{eqnarray*}
\langle B\rangle_{T}=\lim\limits_{\Delta
t_{k}\rightarrow0}\sum\limits_{k=1}^{m-1}|B_{t_{k+1}}-B_{t_{k}}|^{2}.
\end{eqnarray*}

The preceding construction is the canonical continuous-time model for a world in
which investors face ambiguity about drift and volatility.

The set of priors leads naturally to a sublinear expectation:
$$\mathbb{\hat{E}}[\cdot]=\sup\limits_{P\in\mathcal {P}} E_P[\cdot].$$

One advantage of our continuous-time uncertainty model is the fact that one can describe uncertainty by a quadratic real function.
The sublinear function  $G: \mathbb{R}^{d}\times
\mathbb{S}_{d}\rightarrow\mathbb{R}:$
\begin{eqnarray*}
G(p,A)=\sup\limits_{(q,Q)\in\Theta}\Big\{\langle
p,q\rangle+\dfrac{1}{2}tr(AQ)\Big\}, \ \text{for }\
(p,A)\in\mathbb{R}^{d}\times \mathbb{S}_{d},
\end{eqnarray*}
 where $\mathbb{S}_{d}$ is the collection of $d\times d$ symmetric
 matrices, will describe locally, at the level of parameters, uncertainty of
drift and volatility in our model.

 Let  $(B,b)$ be a pair of random vectors under $\mathbb{\hat{E}}$ such that $B$ is a $G$-Brownian motion and
 $b$ is a $G$-distributed process, which just has mean uncertainty. $B$ is a $G$-normal distributed process,
 which just has volatility uncertainty. To see this, we consider
 $d=1$ and
 $\Theta=[\underline{\mu},\overline{\mu}]\times[\underline{\sigma}^{2},\overline{\sigma}^{2}]$.
 Then the process $b$ has mean uncertainty with parameters
 $[\underline{\mu},\overline{\mu}]$, i.e.,
  \begin{eqnarray*}
\mathbb{\hat{E}}[b_{t}]=\overline{\mu}t,\ \mbox{and}\
-\mathbb{\hat{E}}[-b_{t}]=\underline{\mu}t.
\end{eqnarray*}
 And the process $B$ does not have mean uncertainty,
 i.e.,
 \begin{eqnarray*}
\mathbb{\hat{E}}[B_{t}]=\mathbb{\hat{E}}[-B_{t}]=0,
\end{eqnarray*}  but has volatility
 uncertainty with parameters
 $[\underline{\sigma}^{2},\overline{\sigma}^{2}]$, i.e.,
   \begin{eqnarray*}
\mathbb{\hat{E}}[B^{2}_{t}]=\overline{\sigma}^{2}t,\ \mbox{and}\
-\mathbb{\hat{E}}[-B^{2}_{t}]=\underline{\sigma}^{2}t.
\end{eqnarray*}
 The interested reader can refer to the Appendix for the basic
definition and fundamental results of $G$-Brownian motion. The
preceding construction is the canonical model for a world in which
investors face ambiguity about drift and volatility, but do know
certain bounds on these processes.

\subsection{The Canonical Model}

While the abstract characterization of optimal policies holds in a very general setting,
we will frequently focus on the special case where ambiguity about drift is independent of
ambiguity about volatility of the individual asset returns.

For given constants $\underline{\mu}^{i}\leq\overline{\mu}^{i},
\underline{\sigma}^{i}\leq\overline{\sigma}^{i}, i=1,\cdots,d,$ we
consider
$$[\underline{\mu},\overline{\mu}]=\Big\{[\mu^{1},\cdots,\mu^{d}]^{T},
\mu^{i}\in[\underline{\mu}^{i},\overline{\mu}^{i}],
i=1,\cdots,d\Big\},$$ and
$$\Gamma=\Big\{diag[\gamma^{1},\cdots,\gamma^{d}],
\gamma^{i}\in[(\underline{\sigma}^{i})^{2},(\overline{\sigma}^{i})^{2}],
i=1,\cdots,d\Big\}.$$

In order to give an explicit solution, we consider a special case of
$\Theta=[\underline{\mu},\overline{\mu}]\times\Gamma$, and
$A=diag\{1,\cdots,1\}$.  We call this the \emph{canonical} model.

\section{Proofs}

\subsection{Properties of $V(x_{0})$ }

\begin{proposition}\label{mp1}
 $V(x_{0})$  is increasing and concave in  $x$.
\end{proposition}
\noindent {\bf Proof}.  Just for the proof of this proposition, we
denote by
\begin{eqnarray*}
J(\pi,c, x_{0})=\mathbb{\underline{E}}[\int_{0}^{T}u(s,
c_{s})ds+\Phi(T,X_T)],
\end{eqnarray*}
    Also we denote the solution of (\ref{p3}) by  $X^{x_{0}}$. For any
     arbitrary $0< x\leq y $, by the Comparison theorem of
     stochastic differential equations driven by $G$-Brownian motion
     we have $X^{x}\leq X^{y}$. Since the utility function
$u$ and the bequest function $\Phi$  are strictly increasing, by the
the monotonicity of $\underline{\mathbb{E}}$-expectation, we know
that $J$ is increasing in $x$. Therefore, $V$ is increasing  in $x$.

For any arbitrary $0< x_{1},  x_{2} $ and $\lambda\in [0,1]$, we
denote by $x_{\lambda}=\lambda x_{1}+ (1-\lambda)x_{2}$. For any
$c^{1},c^{2}\in C$ and $\pi^{1}, \pi^{2}\in\Pi$, we consider

\begin{eqnarray*}
\left\{\begin{array}{l l}
      dX^{1}_{t}=rX^{1}_{t}(1-(\pi_{t}^{1})^{T}\mathbbold{1})dt+X_{t}^{1}(\pi_{t}^{1})^{T}db_{t}-c^{1}_{t}dt
      +X^{1}_{t}(\pi_{t}^{1})^{T}AdB_{t},\\
      X^{1}_{0}=x_{1},
         \end{array}
  \right.
\end{eqnarray*}
and
\begin{eqnarray*}
\left\{\begin{array}{l l}
      dX^{2}_{t}=rX^{2}_{t}(1-(\pi_{t}^{2})^{T}\mathbbold{1})dt+X_{t}^{2}(\pi_{t}^{2})^{T}db_{t}-c^{2}_{t}dt
      +X^{2}_{t}(\pi_{t}^{2})^{T}AdB_{t},\\
      X^{2}_{0}=x_{2},
         \end{array}
  \right.
\end{eqnarray*}
We denote by $X^{\lambda}:=\lambda X^{1}+ (1-\lambda)X^{2}$,
$c^{\lambda}:=\lambda c^{1}+ (1-\lambda)c^{2}$ and
\begin{eqnarray*}
\pi^{\lambda}:=\frac{\lambda \pi^{1} X^{1}+ (1-\lambda)\pi^{2}
X^{2}}{\lambda X^{1}+ (1-\lambda)X^{2}}.
\end{eqnarray*}
Then $c^{\lambda}\in C$, $\pi^{\lambda}\in\Pi$ and $X^{\lambda}$
satisfies the following
\begin{eqnarray*}
\left\{\begin{array}{l l}
      dX^{\lambda}_{t}=rX^{\lambda}_{t}(1-(\pi_{t}^{\lambda})^{T}\mathbbold{1})dt
      +X_{t}^{\lambda}(\pi_{t}^{\lambda})^{T}db_{t}-c^{\lambda}_{t}dt
      +X^{\lambda}_{t}(\pi_{t}^{\lambda})^{T}AdB_{t},\\
      X^{\lambda}_{0}=x_{\lambda},
         \end{array}
  \right.
\end{eqnarray*}

Since the functions $u$ and  $\Phi$  are  concave with respect to
$c$ and $X$, respectively, we have
\begin{eqnarray*}
\Phi(T, X_T^{\lambda})\geq \lambda\Phi(T, X_T^{1})+
(1-\lambda)\Phi(T, X_T^{2}).
\end{eqnarray*}
and
\begin{eqnarray*}
u(s, X_s^{\lambda})\geq \lambda u(s, X_s^{1})+ (1-\lambda)u(s,
X_s^{2}), s\in [0,T].
\end{eqnarray*}
 Therefore, by virtue of the positive homogeneity and
subadditivity of $\hat{\mathbb{E}}$, the following holds
\begin{eqnarray*}
&&\mathbb{\underline{E}}[\int_{0}^{T}u(s, c^{\lambda}_{s})ds+\Phi(T,
X^{\lambda}_T)]\\&\geq& \mathbb{\underline{E}}[\int_{0}^{T}u(s,
c^{1}_{s})ds+\Phi(T,
X^{1}_T)]+(1-\lambda)\mathbb{\underline{E}}[\int_{0}^{T}u(s,
c^{2}_{s})ds+\Phi(T, X^{2}_T)],
\end{eqnarray*}
i.e.,
\begin{eqnarray*}
J(\pi^{\lambda},c^{\lambda},x^{\lambda})\geq
J(\pi^{1},c^{1},x_{1})+J(\pi^{2},c^{2}, x_{2}).
\end{eqnarray*}
Consequently,
\begin{eqnarray*}
V(x^{\lambda})\geq J(\pi^{1},c^{1},x_{1})+J(\pi^{2},c^{2}, x_{2}).
\end{eqnarray*}
Since the above holds true for  any $c^{1},c^{2}\in C$ and $\pi^{1},
\pi^{2}\in\Pi$, it follows that
\begin{eqnarray*}
V(x^{\lambda})\geq V(x_{1})+V(x_{2}).
\end{eqnarray*}
This means that $V$ is  concave in  $x$. \qquad $\square$\vskip4mm

\subsection{Proof of Theorem \ref{t1}}
\noindent {\bf Proof}. \ \  For any arbitrary $(\pi, c)\in \Pi\times
C$ we let $X$ be a solution of equation (\ref{p3}) associated with
$(\pi, c)$. From It\^{o} formula it follows that
\begin{eqnarray*}
\varphi(T,X_T)&=&
\int_{0}^{T}[\varphi_{t}(t,X_t)+rX_{t}\varphi_{x}(t,X_t)(1-\pi_{t}^{T}\mathbbold{1})-\varphi_{x}(t,X_t)c_{t}]dt
\\&&+\int_{0}^{T}\varphi_{x}(t,X_t)X_{t}\pi_{t}^{T}db_{t}
+\int_{0}^{T}\varphi_{x}(t,X_t)X_{t}\pi_{t}^{T}AdB_{t}\\&&
+\int_{0}^{T}\frac{1}{2}\varphi_{xx}(t,X_t)X^{2}_{t}\langle
A^{T}\pi_{t}\pi_{t}^{T}A,d<B>_{t}\rangle +\varphi(0,x_0).
\end{eqnarray*}
Since for all $x\in \mathbb{R}$, $\varphi(T,x)= \Phi(T,x)$. Then
taking expectation yields
\begin{eqnarray*}
&&\mathbb{\underline{E}}[\Phi(T,X_T)+\int_{0}^{T}u(t, c_t)dt]\\&=&
\mathbb{\underline{E}}[\int_{0}^{T}[\varphi_{t}(t,X_t)+rX_{t}\varphi_{x}(t,X_t)(1-\pi_{t}^{T}\mathbbold{1})
-\varphi_{x}(t,X_t)c_{t}]dt\\&&
+\int_{0}^{T}\varphi_{x}(t,X_t)X_{t}\pi_{t}^{T}db_{t}
+\int_{0}^{T}\frac{1}{2}\varphi_{xx}(t,X_t)X^{2}_{t}\langle\pi_{t}\pi_{t}^{T},d<B>_{t}\rangle\\&&+\int_{0}^{T}U(t,
c_t)dt\Big] +\varphi(0,x_0)
\\&=&
\mathbb{\underline{E}}\Big[\int_{0}^{T}[\varphi_{t}(t,X_t)+rX_{t}\varphi_{x}(t,X_t)(1-\pi_{t}^{T}\mathbbold{1})
-\varphi_{x}(t,X_t)c_{t}]dt+\int_{0}^{T}u(t,
c_t)dt\\&&-\int_{0}^{T}G(-\varphi_{x}(t,X_t)X_{t}\pi_{t},-X^{2}_{t}\varphi_{xx}(t,X_t)A^{T}\pi_{t}\pi^{T}_{t}A)dt
\\&&+\int_{0}^{T}\varphi_{x}(t,X_t)X_{t}\pi_{t}^{T}db_{t}
+\int_{0}^{T}\frac{1}{2}\varphi_{xx}(t,X_t)X^{2}_{t}\langle\pi_{t}\pi_{t}^{T},d<B>_{t}\rangle
\\&&+\int_{0}^{T}G(-\varphi_{x}(t,X_t)X_{t}\pi_{t},-X^{2}_{t}\varphi_{xx}(t,X_t)A^{T}\pi_{t}\pi^{T}_{t}A)dt\Big]
+\varphi(0,x_0).
\end{eqnarray*}

By virtue of equation (\ref{e3}), we obtain that
\begin{eqnarray*}
&&\mathbb{\underline{E}}[\Phi(T,X_T)+\int_{0}^{T}u(t, c_t)dt]
\\&\leq&
\mathbb{\underline{E}}\Big[\int_{0}^{T}\varphi_{x}(t,X_t)X_{t}\pi_{t}^{T}db_{t}
+\int_{0}^{T}\frac{1}{2}\varphi_{xx}(t,X_t)X^{2}_{t}\langle\pi_{t}\pi_{t}^{T},d<B>_{t}\rangle
\\&&+\int_{0}^{T}G(-\varphi_{x}(t,X_t)X_{t}\pi_{t},-X^{2}_{t}\varphi_{xx}(t,X_t)A^{T}\pi_{t}\pi^{T}_{t}A)dt\Big]
+\varphi(0,x_0)\\&=& \varphi(0,x_0).
\end{eqnarray*}
For the last inequality we use the property of $G$-stochastic
calculus,
\begin{eqnarray*}
&&\mathbb{\underline{E}}\Big[\int_{0}^{T}\varphi_{x}(t,X_t)X_{t}\pi_{t}^{T}db_{t}
+\int_{0}^{T}\frac{1}{2}\varphi_{xx}(t,X_t)X^{2}_{t}\langle\pi_{t}\pi_{t}^{T},d<B>_{t}\rangle
\\&&\qquad +\int_{0}^{T}G(-\varphi_{x}(t,X_t)X_{t}\pi_{t},-X^{2}_{t}\varphi_{xx}(t,X_t)A^{T}\pi_{t}\pi^{T}_{t}A)dt\Big]
=0.
\end{eqnarray*}
Consequently,
\begin{eqnarray*}
 V(x_{0})\leq \varphi(0,x_0).
\end{eqnarray*}
Let $(\hat{\pi}, \hat{c})$ satisfy
\begin{eqnarray*}\label{}
      &\sup\limits_{(\pi, c)\in \mathcal {B}\times\mathcal
{A}}\Big\{u(t,c)+\varphi_{t}(t,x)+xr\varphi_{x}(t,x)(1-\pi^{T}\mathbbold{1})-c\varphi_{x}(t,x)]\nonumber
\\&\quad +\inf\limits_{(q,Q)\in \Theta}\{\varphi_{x}(t,x)x\langle\pi,q\rangle
+\dfrac{1}{2}x^{2}\varphi_{xx}(t,x)\langle
A^{T}\pi\pi^{T}A,Q\rangle\}\Big\}\\ &=
\Big\{U(t,\hat{c})+\varphi_{t}(t,x)+xr\varphi_{x}(t,x)(1-(\hat{\pi})^{T}\mathbbold{1})-\hat{c}\varphi_{x}(t,x)]\nonumber
\\&\quad +\inf\limits_{(q,Q)\in \Theta}\{\varphi_{x}(t,x)x\langle\hat{\pi},q\rangle
+\dfrac{1}{2}x^{2}\varphi_{xx}(t,x)\langle
A^{T}\hat{\pi}(\hat{\pi})^{T}A,Q\rangle\}\Big\},
\end{eqnarray*}
and
\begin{eqnarray*}\label{}
\left\{\begin{array}{l l}
      dX_{t}=[rX_{t}(1-\hat{\pi}(t,X_{t})^{T}\mathbbold{1})dt
      +X_{t}\hat{\pi}(t,X_{t})^{T}db_{t}-\hat{c}(t,X_{t})dt+X_{t}\hat{\pi}(t,X_{t})^{T}AdB_{t},\\
      X_{0}=x_{0}.
         \end{array}
  \right.
\end{eqnarray*}
If   $\pi=\hat{\pi},c=\hat{c}$, then from the above proof  it
follows that
\begin{eqnarray*}
 V(x_{0})= \varphi(0,x_0)
\end{eqnarray*}
  Therefore, we have
\begin{eqnarray*}
V(x_{0})&=&\varphi(0,x_0)= \sup\limits_{(\pi, c)\in \Pi\times C}
J(\pi,c,x_{0})=J(\hat{\pi},\hat{c},x_{0}).
\end{eqnarray*}
This complete the proof.  \ \ $\square$\vskip2mm \vskip4mm

\subsection{Proof of Theorem \ref{t2}}
\noindent {\bf Proof}. \ \ From Theorem \ref{t1}, we consider the
following uncertain HJB equation:
\begin{eqnarray*}\label{}
      &\sup\limits_{(\pi, c)\in \mathcal {B}\times\mathcal
{A}}\Big\{u(t,c)+\varphi_{t}(t,x)+xr\varphi_{x}(t,x)(1-\pi^{T}\mathbbold{1})-c\varphi_{x}(t,x)]\nonumber
\\&\quad +\inf\limits_{(\mu,Q)\in  [\underline{\mu},\overline{\mu}]\times \Gamma}\{\varphi_{x}(t,x)x\langle\pi,q\rangle
+\dfrac{1}{2}x^{2}\varphi_{xx}(t,x)\langle
\pi\pi^{T},Q\rangle\}\Big\}=0,
\end{eqnarray*}
with boundary condition
\begin{eqnarray*}
\varphi(T,x)= \Phi(T,x).
\end{eqnarray*}
The above equation can be written as follows:
\begin{eqnarray*}\label{}
      &\sup\limits_{(\pi, c)\in \mathcal {B}\times\mathcal
{A}}\Big\{u(t,c)+\varphi_{t}(t,x)+xr\varphi_{x}(t,x)(1-\pi^{T}\mathbbold{1})-c\varphi_{x}(t,x)]\nonumber
\\&\quad +\inf\limits_{\mu\in  [\underline{\mu},\overline{\mu}]}\{\varphi_{x}(t,x)x\langle\pi,q\rangle\}
+\dfrac{1}{2}x^{2}\inf\limits_{Q\in\Gamma}\{\varphi_{xx}(t,x)\langle
\pi\pi^{T},Q\rangle\}\Big\}=0,
\end{eqnarray*}

If $\varphi_{x}(t,x)>0 $, then the optimal control
$\hat{\mu}^{i}=\underline{\mu}^{i}1_{\{\pi^{i}>
0\}}+\overline{\mu}^{i}1_{\{\pi^{i}\leq 0\}}, i=1,\cdots, d,$ and
\begin{eqnarray*}
&&\inf\limits_{\mu\in [\underline{\mu},\overline{\mu}]}
\Big\{\varphi_{x}(t,x)x\langle\pi,\mu\rangle\Big\}\\ &=&
\varphi_{x}(t,x)x\sum\limits_{i=1}^{d}\pi^{i}\underline{\mu}^{i}1_{\{\pi^{i}>
0\}}+\varphi_{x}(t,x)x\sum\limits_{i=1}^{d}\pi^{i}\overline{\mu}^{i}1_{\{\pi^{i}\leq
0\}}.
\end{eqnarray*}

If $\varphi_{xx}(t,x)<0,$ then
\begin{eqnarray*}
\inf\limits_{Q\in\Gamma}\{\varphi_{xx}(t,x)\langle
\pi\pi^{T},Q\rangle\}=\varphi_{xx}(t,x)\sum\limits_{i=1}^{d}(\pi^{i})^{2}(\overline{\sigma}^{i})^{2}
\end{eqnarray*}
Therefore, we have the following equation:
\begin{eqnarray}\label{e4}
      &&\sup\limits_{(\pi, c)\in \mathcal {B}\times\mathcal
{A}}\Big\{u(t,c)+\varphi_{t}(t,x)+\varphi_{x}(t,x)xr-\varphi_{x}(t,x)c\nonumber\\
&&\quad\quad\quad+\varphi_{x}(t,x)x\sum\limits_{i=1}^{d}\pi^{i}(\underline{\mu}^{i}-r)1_{\{\pi^{i}>
0\}}+\varphi_{x}(t,x)x\sum\limits_{i=1}^{d}\pi^{i}(\overline{\mu}^{i}-r)1_{\{\pi^{i}\leq
0\}}
\nonumber\\&&\quad\quad\quad+\dfrac{1}{2}x^{2}\varphi_{xx}(t,x)\sum\limits_{i=1}^{d}(\pi^{i})^{2}(\overline{\sigma}^{i})^{2}\Big\}=0.
\end{eqnarray}

From the first order condition it follows that
\begin{eqnarray*}\label{}
    u_{c}(t,\hat{c})=\varphi_{x}(t,x).
\end{eqnarray*}

Since $u_{c}(t,c)$ is decreasing with respect to $c$, then its
inverse exists and it is denoted by $v$. Therefore, the optimal
consumption rule is the following
\begin{eqnarray*}\label{}
    \hat{c}=v(\varphi_{x}(t,x)).
\end{eqnarray*}

\begin{lemma}\label{lemma1}
If  $a=\dfrac{1}{2}\overline{\sigma}^{2}x^{2}\varphi_{xx}(t,x)<0$
and $b=\varphi_{x}(t,x)x>0$,
\begin{eqnarray*}\label{}
   f(\pi)=a\pi^{2} +b\pi(\underline{\mu}-r)1_{\{\pi>
0\}}+b\pi(\overline{\mu}-r)1_{\{\pi\leq 0\}},
\end{eqnarray*}
then  $\sup\limits_{\pi}f(\pi)$ has the  the following three cases.

\begin{enumerate} {}
\item[(i)] If $r\leq\underline{\mu}$, then
 \begin{eqnarray*}\label{}
   \sup\limits_{\pi}f(\pi)=f(\hat{\pi})=-\frac{b^{2}(\underline{\mu}-r)^{2}}{4a}
   =-\frac{\varphi^{2}_{x}(t,x)}{\varphi_{xx}(t,x)} \frac{(\underline{\mu}-r)^{2}}{2\overline{\sigma}^{2}},
\end{eqnarray*}
where
\begin{eqnarray*}\label{}
  \hat{\pi}=-\frac{\varphi_{x}(t,x)}{\varphi_{xx}(t,x)x} \frac{\underline{\mu}-r}{\overline{\sigma}^{2}}.
\end{eqnarray*}

\item[(ii)]
If $\underline{\mu}< r<\overline{\mu}$, then

\begin{eqnarray*}\label{}
   \sup\limits_{\pi}f(\pi)=f(\hat{\pi})=f(0)=0,
\end{eqnarray*}
where $\hat{\pi}=0.$

\item[(iii)] If $\overline{\mu}\leq r$, then
\begin{eqnarray*}\label{}
   \sup\limits_{\pi}f(\pi)=f(\hat{\pi})=-\frac{b^{2}(\overline{\mu}-r)^{2}}{4a}
   =-\frac{\varphi^{2}_{x}(t,x)}{\varphi_{xx}(t,x)} \frac{(\overline{\mu}-r)^{2}}{2\overline{\sigma}^{2}},
\end{eqnarray*}
where
\begin{eqnarray*}\label{}
  \hat{\pi}=-\frac{\varphi_{x}(t,x)}{\varphi_{xx}(t,x)x} \frac{\overline{\mu}-r}{\overline{\sigma}^{2}}.
\end{eqnarray*}

\end{enumerate}

\end{lemma}
\noindent {\bf Proof of Lemma \ref{lemma1}}.

\noindent Case I: If $r\leq\underline{\mu}$, then
\begin{eqnarray*}\label{}
   \sup\limits_{\pi<0}f(\pi)=0,
\end{eqnarray*}
and
\begin{eqnarray*}\label{}
   \sup\limits_{\pi\geq0}f(\pi)=f(\bar{\pi})=-\frac{b^{2}(\underline{\mu}-r)^{2}}{4a}
   =-\frac{\varphi^{2}_{x}(t,x)}{\varphi_{xx}(t,x)} \frac{(\underline{\mu}-r)^{2}}{2\overline{\sigma}^{2}},
\end{eqnarray*}
\begin{eqnarray*}\label{}
  \bar{\pi}=-\frac{\varphi_{x}(t,x)}{\varphi_{xx}(t,x)x} \frac{\underline{\mu}-r}{\overline{\sigma}^{2}}.
\end{eqnarray*}
Therefore,
\begin{eqnarray*}\label{}
   \sup\limits_{\pi}f(\pi)=f(\hat{\pi})=-\frac{b^{2}(\underline{\mu}-r)^{2}}{4a}
   =-\frac{\varphi^{2}_{x}(t,x)}{\varphi_{xx}(t,x)} \frac{(\underline{\mu}-r)^{2}}{2\overline{\sigma}^{2}},
\end{eqnarray*}
where
\begin{eqnarray*}\label{}
  \hat{\pi}=-\frac{\varphi_{x}(t,x)}{\varphi_{xx}(t,x)x} \frac{\underline{\mu}-r}{\overline{\sigma}^{2}}.
\end{eqnarray*}

\noindent Case II: If $\underline{\mu}< r<\overline{\mu}$, then
\begin{eqnarray*}\label{}
   \sup\limits_{\pi<0}f(\pi)=f(0)=0,
\end{eqnarray*}
and
\begin{eqnarray*}\label{}
   \sup\limits_{\pi\geq0}f(\pi)=f(0)=0,
\end{eqnarray*}
Therefore,
\begin{eqnarray*}\label{}
   \sup\limits_{\pi}f(\pi)=f(\hat{\pi})=f(0)=0,
\end{eqnarray*}
where $\hat{\pi}=0.$\vskip4mm

\noindent Case III: If  $\overline{\mu}\leq r$, then
\begin{eqnarray*}\label{}
   \sup\limits_{\pi\geq0}f(\pi)=0,
\end{eqnarray*}
and
\begin{eqnarray*}\label{}
   \sup\limits_{\pi<0}f(\pi)=f(\bar{\pi})=-\frac{b^{2}(\overline{\mu}-r)^{2}}{4a}
   =-\frac{\varphi^{2}_{x}(t,x)}{\varphi_{xx}(t,x)} \frac{(\overline{\mu}-r)^{2}}{2\overline{\sigma}^{2}},
\end{eqnarray*}
where
\begin{eqnarray*}\label{}
  \bar{\pi}=-\frac{\varphi_{x}(t,x)}{\varphi_{xx}(t,x)x} \frac{\overline{\mu}-r}{\overline{\sigma}^{2}}.
\end{eqnarray*}
Therefore,
\begin{eqnarray*}\label{}
   \sup\limits_{\pi}f(\pi)=f(\hat{\pi})=-\frac{b^{2}(\overline{\mu}-r)^{2}}{4a}
   =-\frac{\varphi^{2}_{x}(t,x)}{\varphi_{xx}(t,x)} \frac{(\overline{\mu}-r)^{2}}{2\overline{\sigma}^{2}},
\end{eqnarray*}
where
\begin{eqnarray*}\label{}
  \hat{\pi}=-\frac{\varphi_{x}(t,x)}{\varphi_{xx}(t,x)x} \frac{\overline{\mu}-r}{\overline{\sigma}^{2}}.
\end{eqnarray*}
The proof is complete. \ \ $\square$\vskip2mm

We now turn back to our proof. We define
\begin{eqnarray*}\label{}
   &&f(\pi^{i})=\dfrac{1}{2}(\overline{\sigma}^{i})^{2}x^{2}\varphi_{xx}(t,x)(\pi^{i})^{2} +\varphi_{x}(t,x)x\pi^{i}(\underline{\mu}^{i}-r)1_{\{\pi^{i}>
0\}}\\&&\qquad\qquad
+\varphi_{x}(t,x)x\pi^{i}(\overline{\mu}^{i}-r)1_{\{\pi^{i}\leq
0\}},
\end{eqnarray*}
We want to get $
\sum\limits_{i=1}^{d}\sup\limits_{\pi^{i}}f(\pi^{i}).$ We consider
the following cases.

 \noindent Case I:
$r\leq\inf\limits_{i}\underline{\mu}^{i}$.

From the above lemma it follows that
\begin{eqnarray*}\label{}
   \sum\limits_{i=1}^{d}\sup\limits_{\pi}f(\pi^{i})=\sum\limits_{i=1}^{d}f(\hat{\pi}^{i})
   =-\frac{\varphi^{2}_{x}(t,x)}{\varphi_{xx}(t,x)} \sum\limits_{i=1}^{d}\frac{(\underline{\mu}^{i}-r)^{2}}{2(\overline{\sigma}^{2})^{i}},
\end{eqnarray*}
where
\begin{eqnarray*}\label{}
  \hat{\pi}^{i}=-\frac{\varphi_{x}(t,x)}{\varphi_{xx}(t,x)x} \frac{\underline{\mu}^{i}-r}{(\overline{\sigma}^{2})^{i}}.
\end{eqnarray*}

\noindent Case II:  $\sup\limits_{i}\overline{\mu}^{i}\leq r$.

From the above lemma it follows that
\begin{eqnarray*}\label{}
   \sum\limits_{i=1}^{d}\sup\limits_{\pi}f(\pi^{i})=\sum\limits_{i=1}^{d}f(\hat{\pi}^{i})
   =-\frac{\varphi^{2}_{x}(t,x)}{\varphi_{xx}(t,x)} \sum\limits_{i=1}^{d}\frac{(\overline{\mu}^{i}-r)^{2}}{2(\overline{\sigma}^{2})^{i}},
\end{eqnarray*}
where
\begin{eqnarray*}\label{}
  \hat{\pi}^{i}=-\frac{\varphi_{x}(t,x)}{\varphi_{xx}(t,x)x} \frac{\overline{\mu}^{i}-r}{(\overline{\sigma}^{2})^{i}}.
\end{eqnarray*}

\noindent Case III:  $\inf\limits_{i}\underline{\mu}^{i}<
r<\sup\limits_{i}\overline{\mu}^{i}$.

We denote by $\mathbb{A}_{1}=\{i\ |\ \underline{\mu}^{i}\geq r,
i=1,\cdots,d\}, \mathbb{A}_{2}=\{i\ |\ \overline{\mu}^{i}\leq r,
i=1,\cdots,d\}, \mathbb{A}_{3}=\{i\ |\ \underline{\mu}^{i}< r<
\overline{\mu}^{i}, i=1,\cdots,d\}.$ From the above lemma it follows
that
\begin{eqnarray*}\label{}
   \sum\limits_{i=1}^{d}\sup\limits_{\pi}f(\pi^{i})=\sum\limits_{i=1}^{d}f(\hat{\pi}^{i})
   =-\frac{\varphi^{2}_{x}(t,x)}{\varphi_{xx}(t,x)} \sum\limits_{i\in\mathbb{A}_{1} }\frac{(\underline{\mu}^{i}-r)^{2}}{2(\overline{\sigma}^{2})^{i}}
   -\frac{\varphi^{2}_{x}(t,x)}{\varphi_{xx}(t,x)} \sum\limits_{i\in \mathbb{A}_{1}}\frac{(\overline{\mu}^{i}-r)^{2}}{2(\overline{\sigma}^{2})^{i}},
\end{eqnarray*}
where for $i\in \mathbb{A}_{1}$
\begin{eqnarray*}\label{}
  \hat{\pi}^{i}=-\frac{\varphi_{x}(t,x)}{\varphi_{xx}(t,x)x}
  \frac{\underline{\mu}^{i}-r}{(\overline{\sigma}^{2})^{i}};
\end{eqnarray*}
for $i\in \mathbb{A}_{2}$
\begin{eqnarray*}\label{}
  \hat{\pi}^{i}=-\frac{\varphi_{x}(t,x)}{\varphi_{xx}(t,x)x} \frac{\overline{\mu}^{i}-r}{(\overline{\sigma}^{2})^{i}}.
\end{eqnarray*}
for $i\in \mathbb{A}_{3}, \hat{\pi}^{i}=0.$

 (\ref{e4}) is equivalent to the following equation

\begin{eqnarray}\label{e5}
\left\{\begin{array}{l l}
     u(t,v(\varphi_{x}(t,x)))+\varphi_{t}(t,x)+\varphi_{x}(t,x)xr-\varphi_{x}(t,x)v(\varphi_{x}(t,x))
      \\ \quad\quad\quad-\sum\limits_{i}\dfrac{\varphi^{2}_{x}(t,x)(\underline{\mu}^{i}-r)^{2}}{2(\overline{\sigma}^{i})^{2}
      \varphi_{xx}(t,x)}1_{\{r\leq\underline{\mu}^{i} \}}-
\sum\limits_{i}\dfrac{\varphi^{2}_{x}(t,x)(\overline{\mu}^{i}-r)^{2}}{2(\overline{\sigma}^{i})^{2}\varphi_{xx}(t,x)}1_{\{\overline{\mu}^{i}\leq
r\}}=0,\\
      \varphi(T,x)= \Phi(T,x).
         \end{array}
  \right.
\end{eqnarray}
\vskip3mm

\subsection{Proof of Propositions \ref{ep1}}

\noindent {\bf Proof of Proposition \ref{ep1}}. \ \   We just give
the proof of (i), since the proof of (ii) and (iii) are similar. By
the definition of $u$ and $\Phi$, then the equation $(\ref{e5})$ has
the following form
\begin{eqnarray}\label{e6}
\left\{\begin{array}{l l}
     \dfrac{\varphi_{x}^{1-\alpha^{-1}}}{1-\alpha}+\varphi_{t}+\varphi_{x}xr-\varphi_{x}(t,x)v(\varphi_{x})
     -\sum\limits_{i}\dfrac{\varphi^{2}_{x}(t,x)(\underline{\mu}^{i}-r)^{2}}{2(\overline{\sigma}^{i})^{2}\varphi_{xx}}=0,\\
      \varphi(T,x)= \dfrac{Kx^{1-\alpha}}{1-\alpha}.
         \end{array}
  \right.
\end{eqnarray}
We suppose that $\varphi(t,x)$ has the following form
\begin{eqnarray*}\label{}
 \varphi(t,x)=f(t)\dfrac{x^{1-\alpha}}{1-\alpha},
\end{eqnarray*}
where $f(t)$ is a function and given later. Therefore, substituting
the above form of  $\varphi(t,x)$ in to (\ref{e6}),  we obtain the
following equation
\begin{eqnarray*}\label{}
\left\{\begin{array}{l l}
      \alpha f(t)^{^{1-\alpha^{-1}}}  +\beta f(t) +f'(t)=0,\\
      f(T)=K,
\end{array}
  \right.
\end{eqnarray*}
where
\begin{eqnarray*}\label{}
  \beta=\big[r+\sum\limits_{i}\frac{(\underline{\mu}^{i}-r)^{2}}{2(\overline{\sigma}^{i})^{2}\alpha}\big](1-\alpha).
\end{eqnarray*}
The solution of the above equation is given by

\begin{eqnarray*}
f(t)=\Big[K^{\alpha^{-1}}e^{\beta\alpha^{-1}(T-t)}+\alpha
\beta^{-1}(e^{\beta\alpha^{-1}(T-t)}-1)\Big]^{\alpha}.
\end{eqnarray*}
Therefore, from Theorem \ref{t2} we can get the desired result. The
proof is complete. \ \ $\square$

\vskip5mm
\subsection{Proof of Theorem \ref{4t}}

 Similar to the Section 3, we have the
corresponding uncertain HJB as follows.
\begin{eqnarray}\label{e43}
      &\sup\limits_{(\pi, c)\in \mathcal {B}\times\mathcal
{A}}\Big\{u(t,c)+\varphi_{t}(t,x)-c\varphi_{x}(t,x)+\inf\limits_{r\in
[\underline{r},\overline{r}]}\{ xr\varphi_{x}(t,x)(1-\pi)\}
\nonumber
\\&\quad +\inf\limits_{(\mu,\sigma)\in [\underline{\mu},\overline{\mu}]\times[\underline{\sigma},\overline{\sigma}]}
\{\varphi_{x}(t,x)x\pi\mu
+\dfrac{1}{2}x^{2}\varphi_{xx}(t,x)\pi^{2}\sigma^{2}\}\Big\}=0,
\end{eqnarray}
with boundary condition
\begin{eqnarray*}
\varphi(T,x)= \Phi(T,x).
\end{eqnarray*}

Before giving the proof of  Theorem \ref{5t}, we give the following
theorem, which will be needed in what follows.

\begin{theorem}\label{4t}
  Let $\varphi\in C^{1,2}((0,T)\times \mathbb{R}^{+})$  be a
solution of (\ref{e43}) and $\varphi_{xx}< 0$, then the optimal
consumption is
\begin{eqnarray*}\label{}
    \hat{c}=v(\varphi_{x}(t,x)),
\end{eqnarray*}
where $v$ is the inverse of $u_{c}$, and
\begin{enumerate} {}
\item[(i)] if  $\overline{\mu}\leq\underline{r}$, then the optimal portfolio
choice is
\begin{eqnarray*}\label{}
  \hat{\pi}=-\frac{\varphi_{x}(t,x)}{\varphi_{xx}(t,x)x} \frac{\overline{\mu}-\underline{r}}{\overline{\sigma}^{2}}.
\end{eqnarray*}

\item[(ii)] if $\underline{\mu}< \underline{r}<\overline{\mu}$, then
the optimal portfolio choice is $\hat{\pi}=0.$

\item[(iii)]
if $\underline{r}< \underline{\mu}<\overline{r}$, then the optimal
portfolio choice is
 \begin{eqnarray*}\label{}
  \hat{\pi}=\Big[-\frac{\varphi_{x}(t,x)}{\varphi_{xx}(t,x)x} \frac{\underline{\mu}
  -\underline{r}}{\overline{\sigma}^{2}}\Big]\wedge 1.
\end{eqnarray*}

\item[(iv)] if $\underline{\mu}\geq\overline{r}$, and if
 $-\frac{\varphi_{x}(t,x)}{\varphi_{xx}(t,x)} \frac{(\underline{\mu}-\underline{r})}{\overline{\sigma}^{2}}<x$,
 then the optimal portfolio choice is
\begin{eqnarray*}\label{}
  \hat{\pi}=-\frac{\varphi_{x}(t,x)}{\varphi_{xx}(t,x)x}
  \frac{\underline{\mu}-\underline{r}}{\overline{\sigma}^{2}};
 \end{eqnarray*}
 and if $-\frac{\varphi_{x}(t,x)}{\varphi_{xx}(t,x)}
  \frac{(\underline{\mu}-\overline{r})}{\overline{\sigma}^{2}}>x,$
  then then the optimal portfolio choice is
\begin{eqnarray*}\label{}
  \hat{\pi}=-\frac{\varphi_{x}(t,x)}{\varphi_{xx}(t,x)x}
  \frac{\underline{\mu}-\overline{r}}{\overline{\sigma}^{2}};
\end{eqnarray*}
 and if $-\frac{\varphi_{x}(t,x)}{\varphi_{xx}(t,x)}
  \frac{\underline{\mu}-\overline{r}}{\overline{\sigma}^{2}}\leq x\leq
  -\frac{\varphi_{x}(t,x)}{\varphi_{xx}(t,x)} \frac{\underline{\mu}-\underline{r}}{\overline{\sigma}^{2}},$
  then then the optimal portfolio choice is
  $\hat{\pi}=1.$
\end{enumerate}
\end{theorem}

\noindent {\bf Proof of Theorem \ref{4t}}. \ \ From the first order
condition it follows that
\begin{eqnarray*}\label{}
    \hat{c}=v(\varphi_{x}(t,x)).
\end{eqnarray*}
where $v$ is the inverse of $u_{c}(t,c)$.

We denote by
$a=\dfrac{1}{2}\overline{\sigma}^{2}x^{2}\varphi_{xx}(t,x)<0$ and
$b=\varphi_{x}(t,x)x>0$. Let us consider the following functions
\begin{eqnarray*}\label{}
   f(\pi)&=&a\pi^{2} +b\pi(\underline{\mu}-\overline{r})1_{\{\pi>
1\}}+b\pi(\underline{\mu}-\underline{r})1_{\{0\leq\pi\leq 1\}}\\&&\
+b\pi(\overline{\mu}-\underline{r})1_{\{\pi\leq
0\}}+b\overline{r}1_{\{\pi> 1\}}+b\underline{r}1_{\{\pi\leq 1\}}.
\end{eqnarray*}
For this, we define the following functions:
\begin{eqnarray*}\label{}
   f_{1}(\pi)&=&a\pi^{2}
   +b\pi(\underline{\mu}-\overline{r})+b\overline{r},\ \pi>
1, \\
f_{2}(\pi)&=&a\pi^{2}
+b\pi(\underline{\mu}-\underline{r})+b\underline{r},\ 0\leq\pi\leq 1,\\
f_{3}(\pi)&=&a\pi^{2}
+b\pi(\overline{\mu}-\underline{r})+b\underline{r},\ \pi\leq 0.
\end{eqnarray*}
 Let us consider
$\sup\limits_{\pi}f(\pi)$ in the following cases.

\noindent Case I: If $\overline{\mu}\leq\underline{r}$, then
\begin{eqnarray*}\label{}
   \sup\limits_{\pi>1}f_{1}(\pi)=f_{1}(1)=a +b\underline{\mu},
\end{eqnarray*}
\begin{eqnarray*}\label{}
   \sup\limits_{0\leq\pi\leq 1}f_{2}(\pi)=f_{2}(0)=b\underline{r},
\end{eqnarray*}
\begin{eqnarray*}\label{}
   \sup\limits_{\pi< 0}f_{3}(\pi)=f_{3}(\overline{\pi})=b\underline{r}-\dfrac{b^{2}(\overline{\mu}-\underline{r})^{2}}{4a},
\end{eqnarray*}
where
\begin{eqnarray*}\label{}
  \overline{\pi}=-\frac{\varphi_{x}(t,x)}{\varphi_{xx}(t,x)x} \frac{\overline{\mu}-\underline{r}}{\overline{\sigma}^{2}}.
\end{eqnarray*}
Since $a<0$ and $b>0$, we have
\begin{eqnarray*}\label{}
   \sup\limits_{\pi}f(\pi)=f(\hat{\pi}),
\end{eqnarray*}
where
\begin{eqnarray*}\label{}
  \hat{\pi}=-\frac{\varphi_{x}(t,x)}{\varphi_{xx}(t,x)x} \frac{\overline{\mu}-\underline{r}}{\overline{\sigma}^{2}}.
\end{eqnarray*}

\noindent Case II: If $\underline{\mu}<
\underline{r}<\overline{\mu}$, then it follows that
\begin{eqnarray*}\label{}
   \sup\limits_{\pi>1}f_{1}(\pi)=f_{1}(1)=a +b\underline{\mu},
\end{eqnarray*}
\begin{eqnarray*}\label{}
   \sup\limits_{0\leq\pi\leq 1}f_{2}(\pi)=f_{2}(0)=b\underline{r},
\end{eqnarray*}
\begin{eqnarray*}\label{}
   \sup\limits_{\pi< 0}f(\pi)=f_{3}(0)=b\underline{r}.
\end{eqnarray*}
Since $a<0$ and $b>0$, we have
\begin{eqnarray*}\label{}
   \sup\limits_{\pi}f(\pi)=f(\hat{\pi}),
\end{eqnarray*}
where $\hat{\pi}=0.$\vskip4mm

\noindent Case III:  If  $\underline{r}<
\underline{\mu}<\overline{r}$ then
\begin{eqnarray*}\label{}
   \sup\limits_{\pi>1}f_{1}(\pi)=f_{1}(1)=a +b\underline{\mu},
\end{eqnarray*}
\begin{eqnarray*}\label{}
   \sup\limits_{0\leq\pi\leq
   1}f_{2}(\pi)=f_{2}(\bar{\pi})>f_{2}(0)=f_{3}(0), f_{2}(\bar{\pi})>f_{2}(1)=f_{1}(1)
\end{eqnarray*}
where
\begin{eqnarray*}\label{}
  \bar{\pi}=[-\frac{\varphi_{x}(t,x)}{\varphi_{xx}(t,x)x}
   \frac{\underline{\mu}-\underline{r}}{\overline{\sigma}^{2}}]\wedge
   1,
\end{eqnarray*}
\begin{eqnarray*}\label{}
   \sup\limits_{\pi< 0}f_{3}(\pi)=f_{3}(0)=b\underline{r},
\end{eqnarray*}
Therefore, it follows that
\begin{eqnarray*}\label{}
   \sup\limits_{\pi}f(\pi)=f(\hat{\pi}),
\end{eqnarray*}
where
\begin{eqnarray*}\label{}
  \hat{\pi}=[-\frac{\varphi_{x}(t,x)}{\varphi_{xx}(t,x)x} \frac{\underline{\mu}-\underline{r}}{\overline{\sigma}^{2}}]\wedge 1.
\end{eqnarray*}

\noindent Case IV:  $\underline{\mu}\geq\overline{r}$. (a) If
$-\frac{\varphi_{x}(t,x)}{\varphi_{xx}(t,x)}
\frac{\underline{\mu}-\underline{r}}{\overline{\sigma}^{2}}<x$, then
\begin{eqnarray*}\label{}
   \sup\limits_{\pi>1}f_{1}(\pi)=f_{1}(1)=a +b\underline{\mu},
\end{eqnarray*}
\begin{eqnarray*}\label{}
   \sup\limits_{0\leq\pi\leq 1}f_{2}(\pi)=f_{2}(\bar{\pi})>f_{2}(0)=f_{3}(0),
   f_{2}(\bar{\pi})>f_{2}(1)=f_{1}(1),
\end{eqnarray*}
where
\begin{eqnarray*}\label{}
  \bar{\pi}=[\frac{\varphi_{x}(t,x)}{\varphi_{xx}(t,x)x}
   \frac{(\underline{\mu}-\underline{r})}{\overline{\sigma}^{2}},
\end{eqnarray*}
\begin{eqnarray*}\label{}
   \sup\limits_{\pi< 0}f_{3}(\pi)=f_{3}(0)=b\underline{r}.
\end{eqnarray*}
Consequently,
\begin{eqnarray*}\label{}
   \sup\limits_{\pi}f(\pi)=f(\hat{\pi}),
\end{eqnarray*}
and the optimal portfolio choice is
\begin{eqnarray*}\label{}
  \hat{\pi}=-\frac{\varphi_{x}(t,x)}{\varphi_{xx}(t,x)x}
  \frac{(\underline{\mu}-\underline{r})}{\overline{\sigma}^{2}}.
 \end{eqnarray*}
 (b) If $-\frac{\varphi_{x}(t,x)}{\varphi_{xx}(t,x)}
  \frac{\underline{\mu}-\overline{r}}{\overline{\sigma}^{2}}>x,$
  then
\begin{eqnarray*}\label{}
   \sup\limits_{\pi\geq 1}f_{1}(\pi)=f_{1}(\bar{\pi})>f_{2}(0)=f_{3}(0),
   f_{2}(\bar{\pi})>f_{2}(1)=f_{1}(1),
\end{eqnarray*}
where
\begin{eqnarray*}\label{}
  \bar{\pi}=-\frac{\varphi_{x}(t,x)}{\varphi_{xx}(t,x)x}
  \frac{\underline{\mu}-\overline{r}}{\overline{\sigma}^{2}},
\end{eqnarray*}
\begin{eqnarray*}\label{}
   \sup\limits_{0\leq\pi\leq1}f_{2}(\pi)=f_{2}(1)=f_{1}(1)=a
   +b\underline{\mu}>f_{2}(0)=f_{3}(0),
\end{eqnarray*}
\begin{eqnarray*}\label{}
   \sup\limits_{\pi< 0}f_{3}(\pi)=f_{3}(0)=b\underline{r}.
\end{eqnarray*}
From the above it follows that the optimal portfolio choice is
\begin{eqnarray*}\label{}
  \hat{\pi}=-\frac{\varphi_{x}(t,x)}{\varphi_{xx}(t,x)x}
  \frac{\underline{\mu}-\overline{r}}{\overline{\sigma}^{2}}.
\end{eqnarray*}
(c) If $-\frac{\varphi_{x}(t,x)}{\varphi_{xx}(t,x)}
  \frac{\underline{\mu}-\overline{r}}{\overline{\sigma}^{2}}\leq x\leq
  -\frac{\varphi_{x}(t,x)}{\varphi_{xx}(t,x)}
  \frac{\underline{\mu}-\underline{r}}{\overline{\sigma}^{2}},$ then
  \begin{eqnarray*}\label{}
   \sup\limits_{\pi\geq 1}f_{1}(\pi)=f_{1}(1),
\end{eqnarray*}
\begin{eqnarray*}\label{}
   \sup\limits_{0\leq\pi\leq1}f_{2}(\pi)=f_{2}(1)=f_{1}(1)=a
   +b\underline{\mu}>f_{2}(0)=f_{3}(0),
\end{eqnarray*}
\begin{eqnarray*}\label{}
   \sup\limits_{\pi< 0}f_{3}(\pi)=f_{3}(0)=b\underline{r}.
\end{eqnarray*}
Therefore  the optimal portfolio choice is
  $\hat{\pi}=1.$ The
proof is complete. \ \ $\square$\vskip4mm

\vskip5mm

\noindent {\bf Proof of Theorem \ref{5t}.} \vskip2mm

Similar to the proof of Proposition  \ref{ep1}, from Theorem
\ref{4t} we can get that

\begin{enumerate} {}
\item[(i)] if  $\overline{\mu}\leq\underline{r}$, then the optimal portfolio
choice is
\begin{eqnarray*}\label{}
  \hat{\pi}=\frac{\overline{\mu}-\underline{r}}{\alpha\overline{\sigma}^{2}}.
\end{eqnarray*}

\item[(ii)] if $\underline{\mu}< \underline{r}<\overline{\mu}$, then
the optimal portfolio choice is $\hat{\pi}=0.$

\item[(iii)]
if $\underline{r}< \underline{\mu}<\overline{r}$, then the optimal
portfolio choice is
 \begin{eqnarray*}\label{}
  \hat{\pi}=\Big[ \frac{\underline{\mu}
  -\underline{r}}{\alpha\overline{\sigma}^{2}}\Big]\wedge 1.
\end{eqnarray*}
\end{enumerate}

We now consider  $\underline{\mu}\geq\overline{r}$ in the following
cases.

\noindent Case I.  Suppose  $
\dfrac{\underline{\mu}-\underline{r}}{\alpha\overline{\sigma}^{2}}<1$,
from Theorem \ref{4t}, then the equation (\ref{e43}) has the
following form
\begin{eqnarray}\label{4e6}
\left\{\begin{array}{l l}
     \dfrac{\varphi_{x}^{1-\alpha^{-1}}}{1-\alpha}+\varphi_{t}+\varphi_{x}xr-\varphi_{x}v(\varphi_{x})
     -\dfrac{\varphi^{2}_{x}(\underline{\mu}-\underline{r})^{2}}{2\overline{\sigma}^{2}\varphi_{xx}}=0,\\
      \varphi(T,x)= \dfrac{Kx^{1-\alpha}}{1-\alpha}.
         \end{array}
  \right.
\end{eqnarray}
Similar to the proof of Proposition  \ref{ep1}, the optimal
portfolio choice is
\begin{eqnarray*}\label{}
  \hat{\pi}=-\frac{\varphi_{x}(t,x)}{\varphi_{xx}(t,x)x}
  \frac{\underline{\mu}-\underline{r}}{\overline{\sigma}^{2}}=
  \frac{\underline{\mu}-\underline{r}}{\alpha\overline{\sigma}^{2}}.
 \end{eqnarray*}

 \noindent Case II.  Suppose  $
\dfrac{\underline{\mu}-\overline{r}}{\alpha\overline{\sigma}^{2}}>1$,
from Theorem \ref{4t}, then the equation (\ref{e43}) has the
following form
\begin{eqnarray}\label{4e6}
\left\{\begin{array}{l l}
     \dfrac{\varphi_{x}^{1-\alpha^{-1}}}{1-\alpha}+\varphi_{t}+\varphi_{x}xr-\varphi_{x}v(\varphi_{x})
     -\dfrac{\varphi^{2}_{x}(\underline{\mu}-\overline{r})^{2}}{2\overline{\sigma}^{2}\varphi_{xx}}=0,\\
      \varphi(T,x)= \dfrac{Kx^{1-\alpha}}{1-\alpha}.
         \end{array}
  \right.
\end{eqnarray}
Similar to the proof of Proposition  \ref{ep1}, the optimal
portfolio choice is
\begin{eqnarray*}\label{}
  \hat{\pi}=-\frac{\varphi_{x}(t,x)}{\varphi_{xx}(t,x)x}
  \frac{(\underline{\mu}-\overline{r})}{\overline{\sigma}^{2}}=
  \frac{\underline{\mu}-\overline{r}}{\alpha\overline{\sigma}^{2}}.
 \end{eqnarray*}

 \noindent Case II. If
$
\dfrac{\underline{\mu}-\overline{r}}{\alpha\overline{\sigma}^{2}}\leq
1\leq
  \dfrac{\underline{\mu}-\underline{r}}{\alpha\overline{\sigma}^{2}},$
  then then the optimal portfolio choice is
  $\hat{\pi}=1.$  The
proof is complete. \ \ $\square$\vskip4mm

\ifx\undefined\BySame
\newcommand{\BySame}{\leavevmode\rule[.5ex]{3em}{.5pt}\ }
\fi
\ifx\undefined\textsc
\newcommand{\textsc}[1]{{\sc #1}}
\newcommand{\emph}[1]{{\em #1\/}}
\let\tmpsmall\small
\renewcommand{\small}{\tmpsmall\sc}
\fi

\end{document}